\documentclass{spie}%
\usepackage{amsfonts}
\usepackage{amsmath}
\usepackage{amssymb}
\usepackage{graphicx}%
\newtheorem{corollary}[theorem]{Corollary}

\begin{document}

\title{Topological Quantum Computing and the Jones Polynomial}
\author{Samuel J. Lomonaco, Jr.\supit{a} and Louis H. Kauffman\supit{b}
\skiplinehalf\supit{a}Department of Computer Science and Electrical
Engineering, University of Maryland Baltimore County (UMBC), Baltimore,
Maryland 21250, USA\\\supit{b}Department of Mathematics, Statistics, and Computer Science,
University of Illinois at Chicago (UIC), Chicago, Illinois 60607, USA }
\authorinfo{Send correspondence to S.J.L: E-mail: Lomonaco@umbc.edu}
\maketitle

\begin{abstract}
In this paper, we give a description of a recent quantum algorithm created by
Aharonov, Jones, and Landau for approximating the values of the Jones
polynomial at roots of unity of the form $e^{2\pi i/k}$. This description is
given with two objectives in mind. The first is to describe the algorithm in
such a way as to make explicit the underlying and inherent control structure.
The second is to make this algorithm accessible to a larger audience.

\end{abstract}
\tableofcontents

\section{Introduction}

\bigskip

In this paper, we give a description of a recent quantum algorithm created by
Aharonov, Jones, and Landau \cite{Aharonov1} for approximating the values of
the Jones polynomial at roots of unity of the form $e^{2\pi i/k}$, for $k$ a
positive integer. We do so with two objectives in mind. The first is to
describe the algorithm in such a way as to make explicit the underlying and
inherent control structure. The second is to make this algorithm accessible to
a larger audience.

\bigskip

To avoid cluttering this exposition, we focus solely on the version of this
quantum algorithm based on the Markov trace closure of a braid. \ The
alternative platt closure version is left as an exercise for the reader.

\bigskip

Readers already familiar with knot theory may want to skip to section 7.

\bigskip

\section{Knot theory is ...}

\bigskip

In its most general form, knot theory is the study of the fundamental problem
of placement:

\bigskip

\noindent\textbf{The Placement Problem.} \ \textit{When are two placements of
a space }$X$\textit{ in a space }$Y$\textit{ the same or different?}%

\begin{center}
\includegraphics[
trim=0.000000in 1.003446in 0.000000in 0.000000in,
natheight=7.499600in,
natwidth=9.999800in,
height=1.9778in,
width=3.0277in
]%
{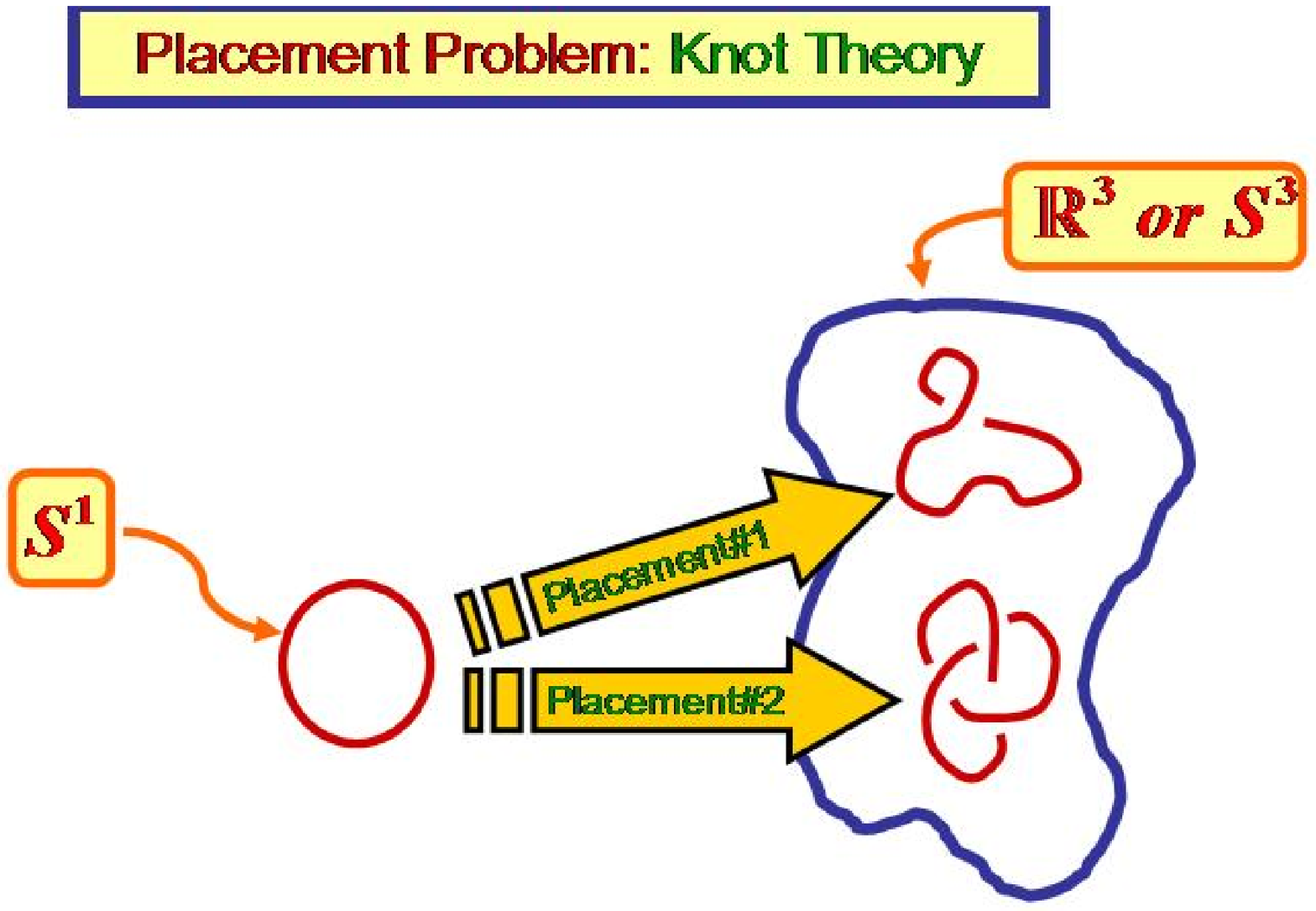}%
\\
\textbf{Figure 1. \ The Placement Problem.}%
\end{center}

In its most renowned form, knot theory is the study of the placement of a
1-sphere\footnote{By 1-sphere we mean a circle.} $S^{1}$(or a disjoint union
of 1-spheres) in real 3-space $\mathbb{R}^{3}$ (or the 3-sphere $S^{3}$),
called the \textbf{ambient space}. \ In this case, "placement" usually means a
smooth (or piecewise linear) embedding, i.e., a smooth homeomorphism into the
ambient space. \ Such a placement is called a \textbf{knot} if a single
1-sphere is embedded ( or a \textbf{link}, if a disjoint union of many
1-spheres is embedded.)

\bigskip

Two knots (or links) are said to be the same, i.e., of the \textbf{same knot
type}, if there exists an orientation preserving autohomeomorphism\footnote{A
provenly equivalent definition is that two knots are of the same knot type if
and only if there exists an isotopy of the ambient space that carries one knot
onto the other.} of the ambient space carrying one knot into the other.
Otherwise, they are said to be different, i.e., of \textbf{different knot
type}. \ Such knots are frequently represented by a knot diagram, i.e., a
planar 4-valent graph with vertices appropriately labelled as
undercrossings/overcrossings, as shown in figure 2.

\bigskip%

\begin{center}
\includegraphics[
trim=0.000000in 2.006893in 0.000000in 0.000000in,
natheight=7.499600in,
natwidth=9.999800in,
height=1.6751in,
width=3.0277in
]%
{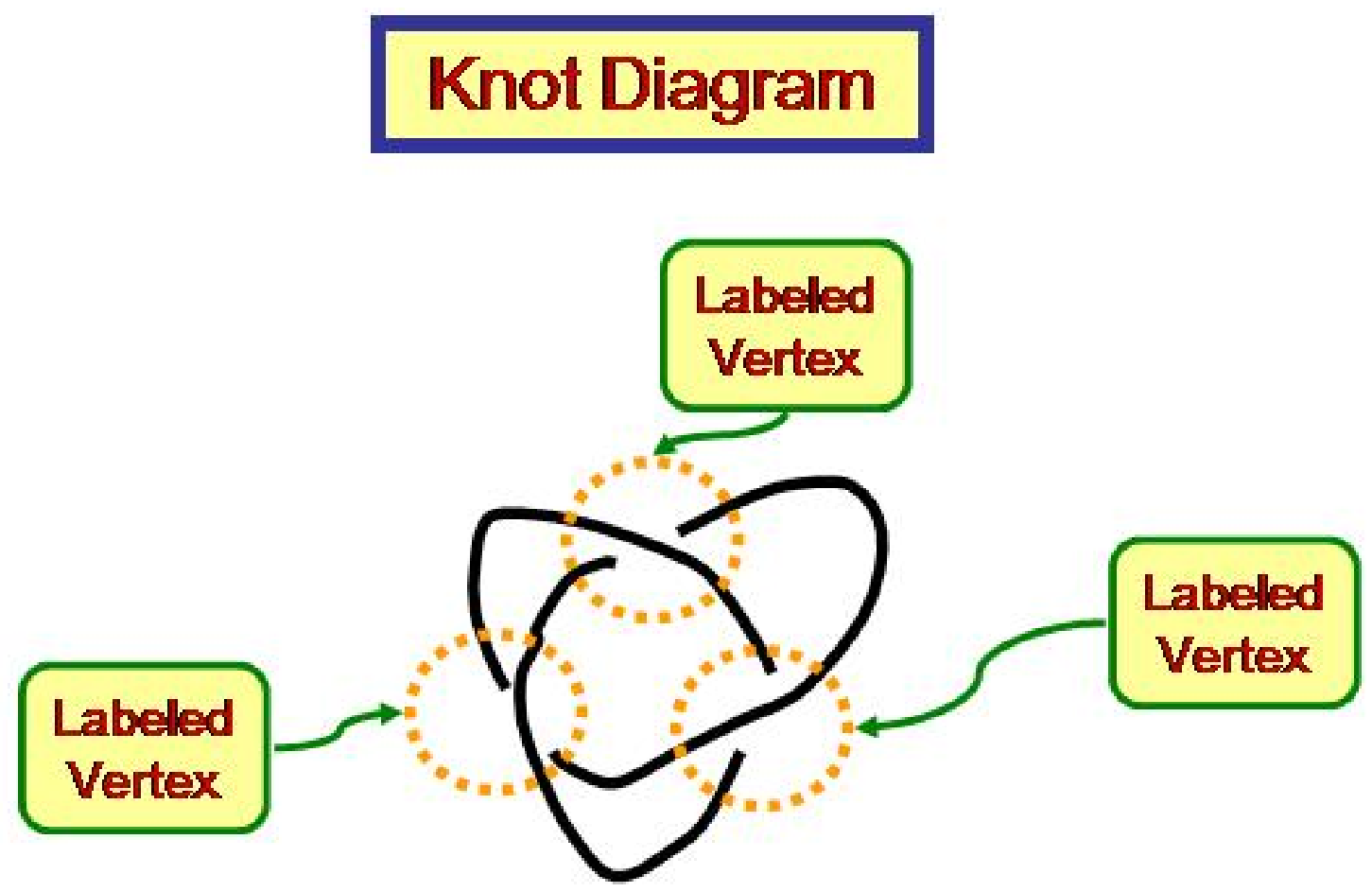}%
\\
\textbf{Figure 2. \ A knot diagram.}%
\end{center}

\bigskip

The fundamental problem of knot theory can now be stated as:

\bigskip

\noindent\textbf{The Fundamental Problem of Knot Theory.} \ \textit{When are
two knots of the same or of different knot type?}

\bigskip

A useful knot theoretic research tool is Reidemeister's theorem, which makes
use of the Reidemeister moves as defined in figure 3.:

\bigskip

\begin{theorem}
[\textbf{Reidemeister}]Two knot diagrams represent the same knot type if and
only if it is possible to transform one into the other by applying a finite
sequence of Reidemeister moves.
\end{theorem}

\textbf{%
\begin{center}
\includegraphics[
natheight=7.499600in,
natwidth=9.999800in,
height=2.2779in,
width=3.0277in
]%
{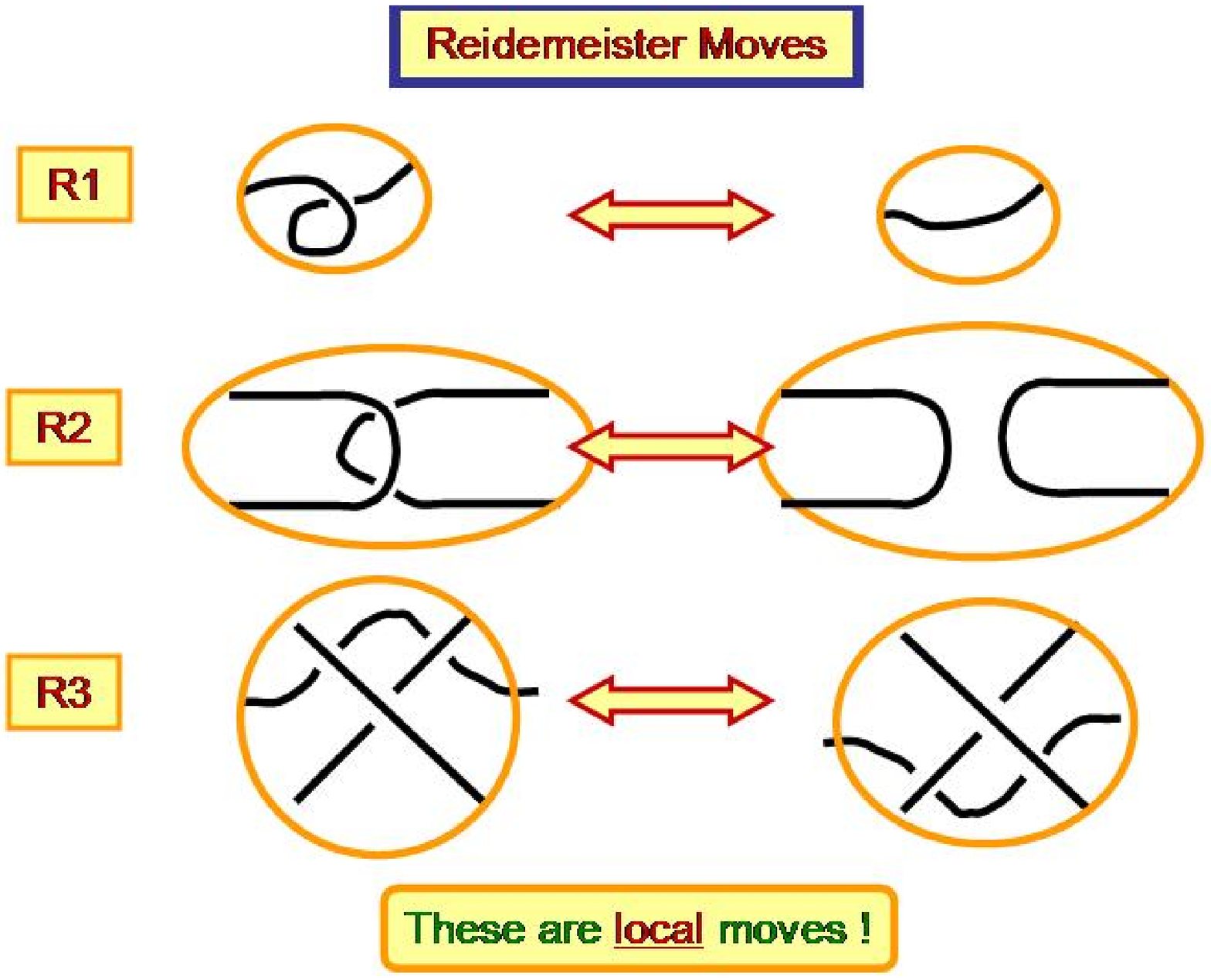}%
\\
\textbf{Figure 3. The Reidemeister moves.}%
\end{center}
}

\bigskip

The standard approach to attacking the fundamental problem of knot theory is
to create knot invariants for distinquishing knots. \ By a \textbf{knot
invariant} $I$ we mean a map from knots to a specified mathematical domain
which maps knots (or links) of the same type to the same mathematical object.
\ \textit{Thus, if an invariant is found to be different on two knots (or
links), then the two knots (or links) cannot be of the same knot type}!

\bigskip

The Jones polynomial is one such invariant, and indeed a very significant one.
\ The Jones polynomial maps each knot (or link) to a Laurent polynomial with
integer coefficients, i.e., an invariant with domain the Laurent ring
$\mathbb{Z}\left[  t,t^{-1}\right]  $. \ If two knots (or links) have
different Jones polynomials then they must be different. \ A definition of
this famous knot invariant is given in section 6 of this paper.

\bigskip

But before we define the Jones polynomial we will need to look at a group
invented by Emil Artin, i.e., the braid group, which is now beginning to have
an impact on quantum computing.

\bigskip

\section{The braid group $B_{n}$}

\bigskip

\begin{definition}
The $n$\textbf{-stranded braid group} $B_{n}$ is is the group generated by the
symbols $b_{1},b_{2},\ldots,b_{n-1}$ subject to the defining relations
\[
\left\{
\begin{array}
[c]{ll}%
b_{i}b_{i+1}b_{i}=b_{i+1}b_{i}b_{i+1} & \text{for \ }1\leq i<n\\
& \\
b_{i}b_{j}=b_{j}b_{i} & \text{for \ }\left\vert i-j\right\vert \geq2
\end{array}
\right.
\]

\end{definition}

The braid group $B_{n}$ is easily understood in terms of diagrammatics, as
shown in Figures 4 to 13.

\bigskip

As illustrated in Figure 4, an element of the braid group $B_{3}$ can be
thought of as a hatbox with the three points at the top connected to the three
points at the bottom by smooth nonintersecting curves, called \textbf{strands}%
. \ As illustrated in Figures 5 and 6, two such elements of $B_{n}$ are equal
if it is possible to continuously move within the hatbox the strands of one
braid into the strands of the other without cutting or breaking the strands
and without letting the strands pass through one another. \ Figure 5 shows two
equal braids, and Figure 6 shows two braids that cannot be equal. \ As
illustrated in Figure 7, the enclosing hatbox is usually omitted, but
understood to be there.

\bigskip

The product of two elements of $B_{n}$ is defined, as illustrated in Figure 8,
by simply stacking one hatbox on top of the other. \ Under this definition of
multiplication, it can be shown that every braid has a multiplicative inverse.
An example is given in Figure 9 of the inverse of a braid. \ The braid group
actually has a finite set of generators, as shown in Figure 10. \ A complete
set of defining relations among these generators are shown in Figures 11 and 12.

\bigskip

\qquad%
{\parbox[b]{2.5278in}{\begin{center}
\includegraphics[
natheight=7.499600in,
natwidth=9.999800in,
height=1.9026in,
width=2.5278in
]%
{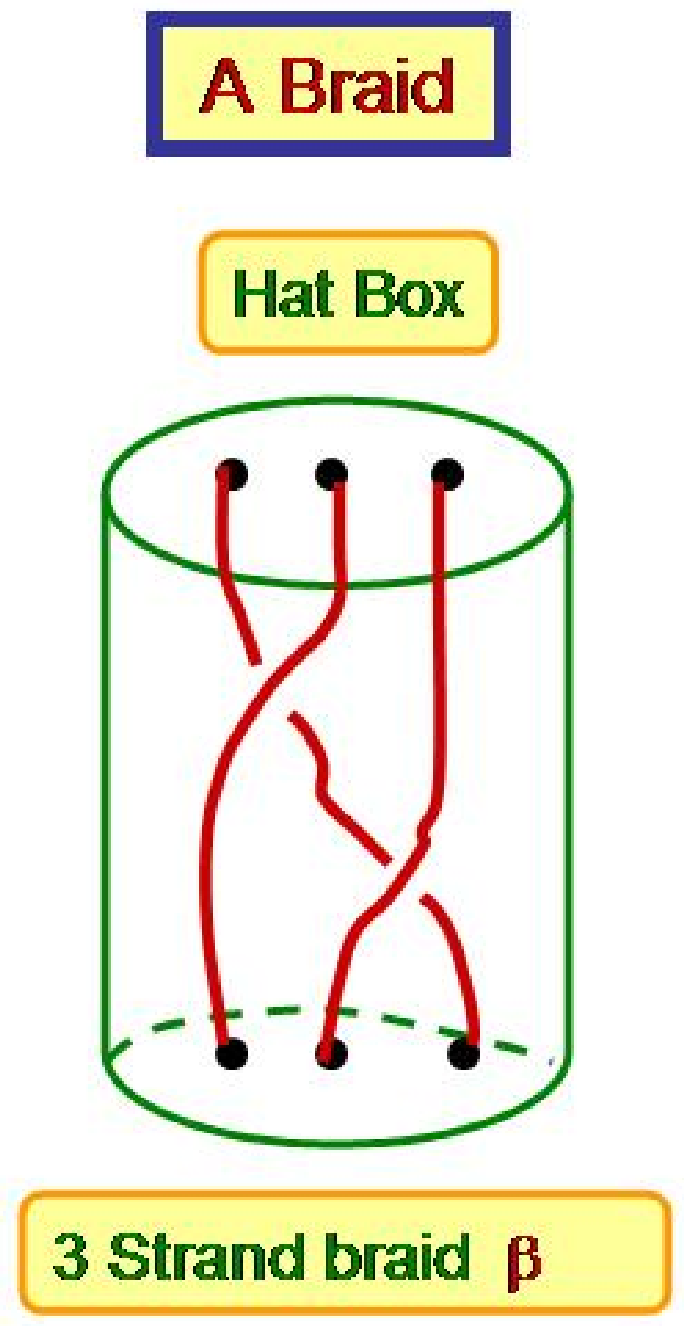}%
\\
\textbf{Figure 4. \ An example of a three stranded braid in }$B_{3}$.
\end{center}}}%
\qquad\qquad%
{\parbox[b]{2.5278in}{\begin{center}
\includegraphics[
natheight=7.499600in,
natwidth=9.999800in,
height=1.9026in,
width=2.5278in
]%
{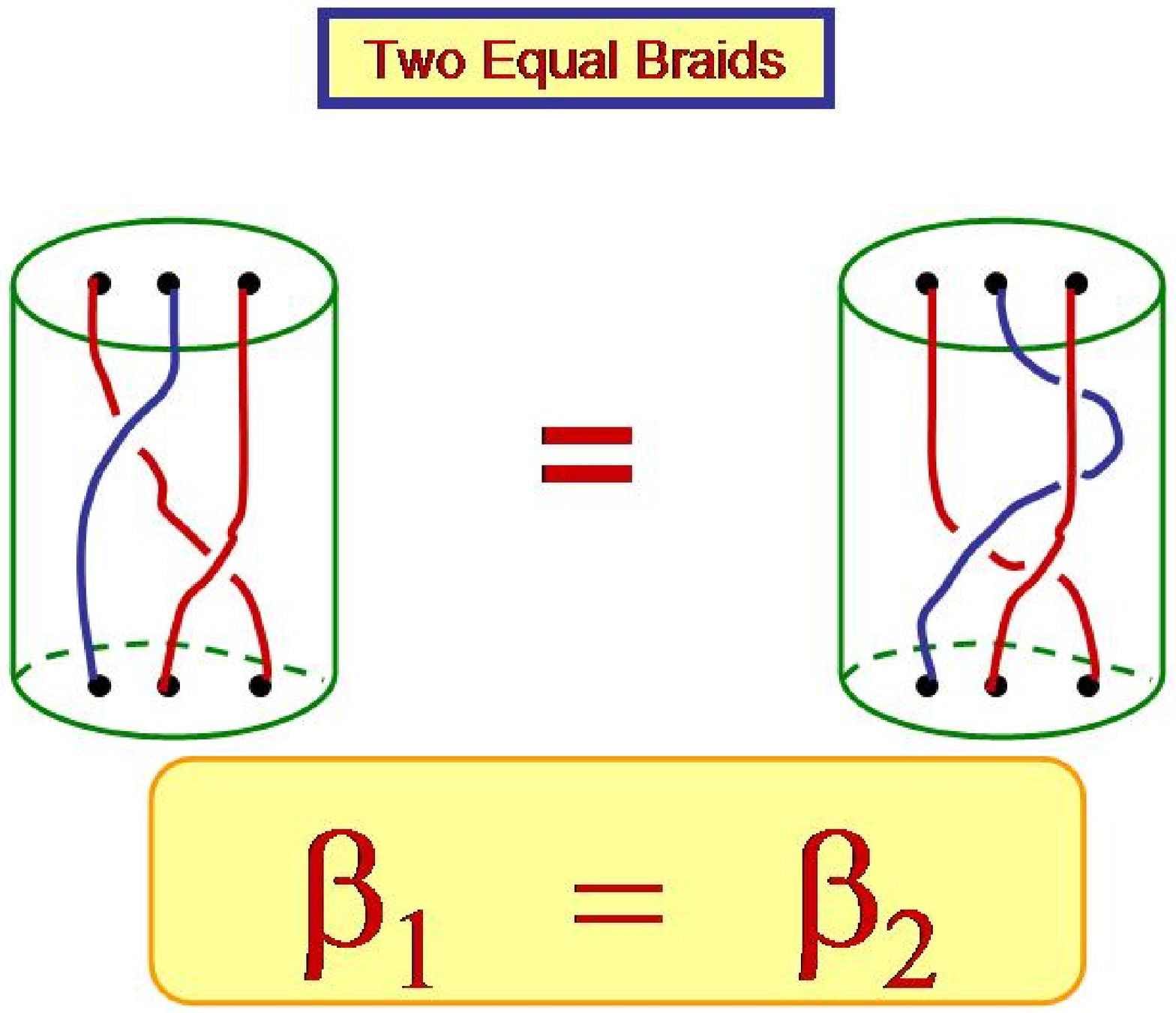}%
\\
\textbf{Figure 5. An example of two equal braids.}%
\end{center}}}%

\qquad%
{\parbox[b]{2.5278in}{\begin{center}
\includegraphics[
natheight=7.499600in,
natwidth=9.999800in,
height=1.9026in,
width=2.5278in
]%
{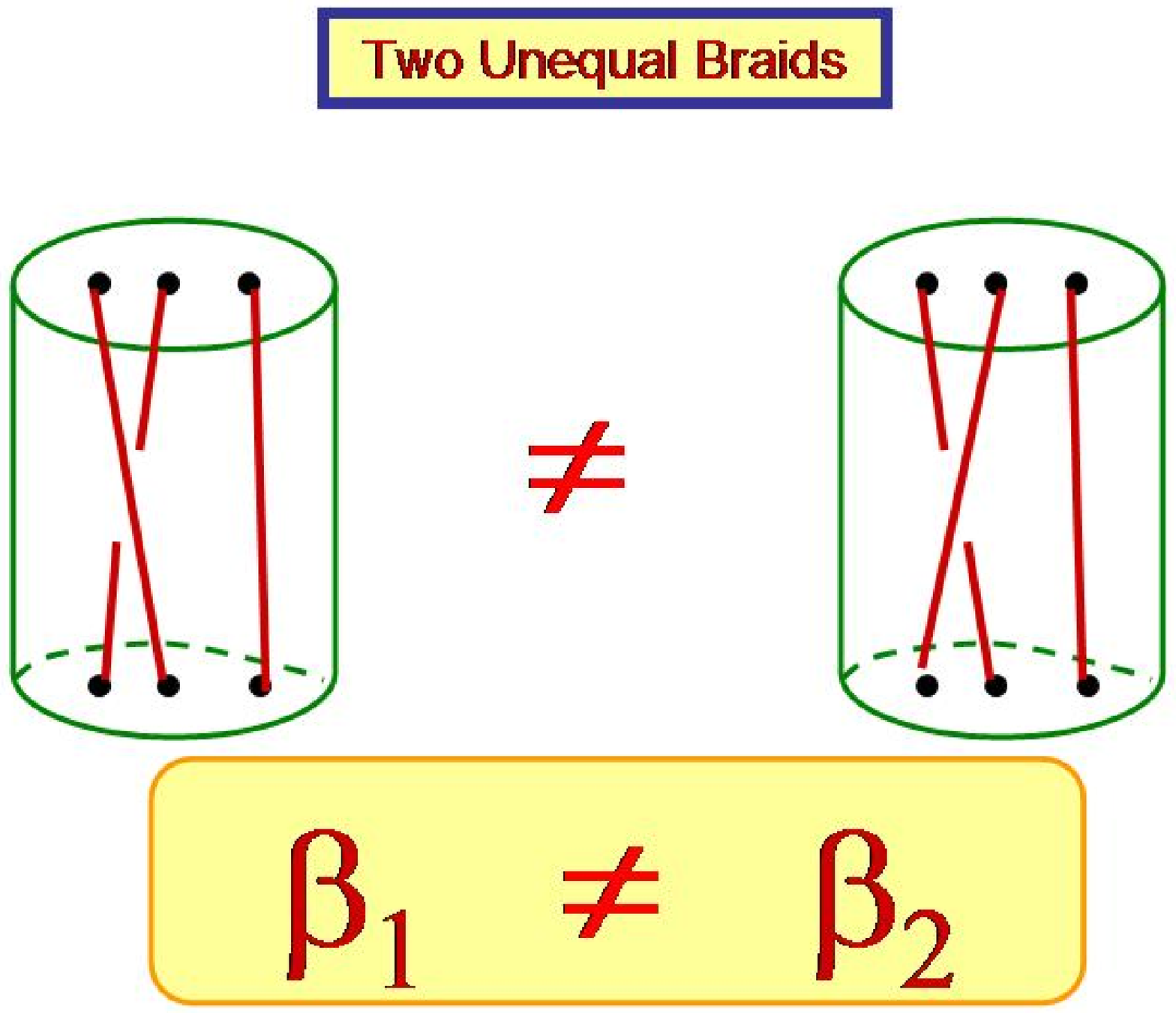}%
\\
\textbf{Figure 6. \ An example of two non-equal braids.}%
\end{center}}}%
\qquad\qquad%
{\parbox[b]{2.5278in}{\begin{center}
\includegraphics[
natheight=7.499600in,
natwidth=9.999800in,
height=1.9026in,
width=2.5278in
]%
{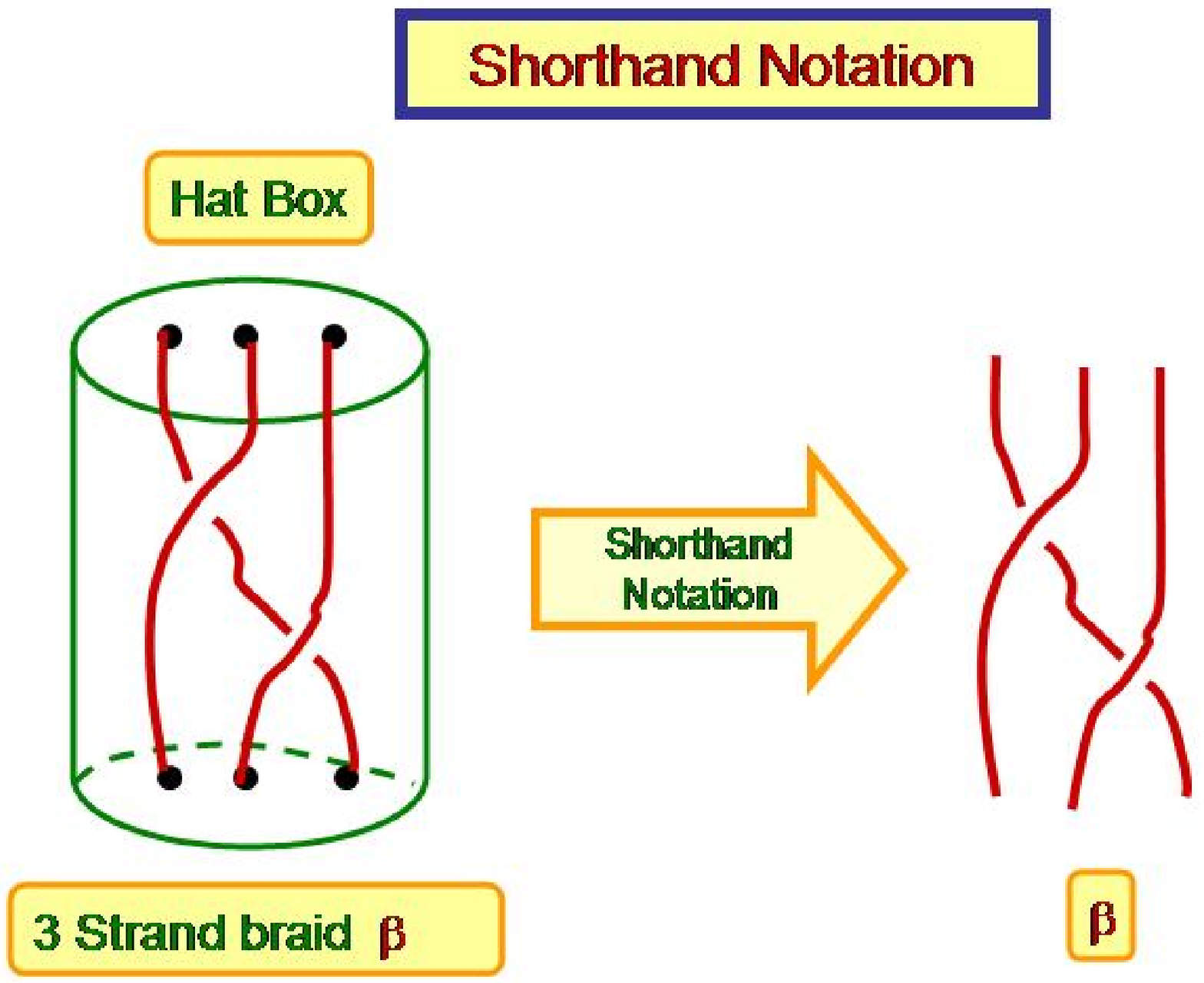}%
\\
\textbf{Figure 7. Shorthand notation for braids.}%
\end{center}}}%

\qquad%
{\parbox[b]{2.5278in}{\begin{center}
\includegraphics[
natheight=7.499600in,
natwidth=9.999800in,
height=1.9026in,
width=2.5278in
]%
{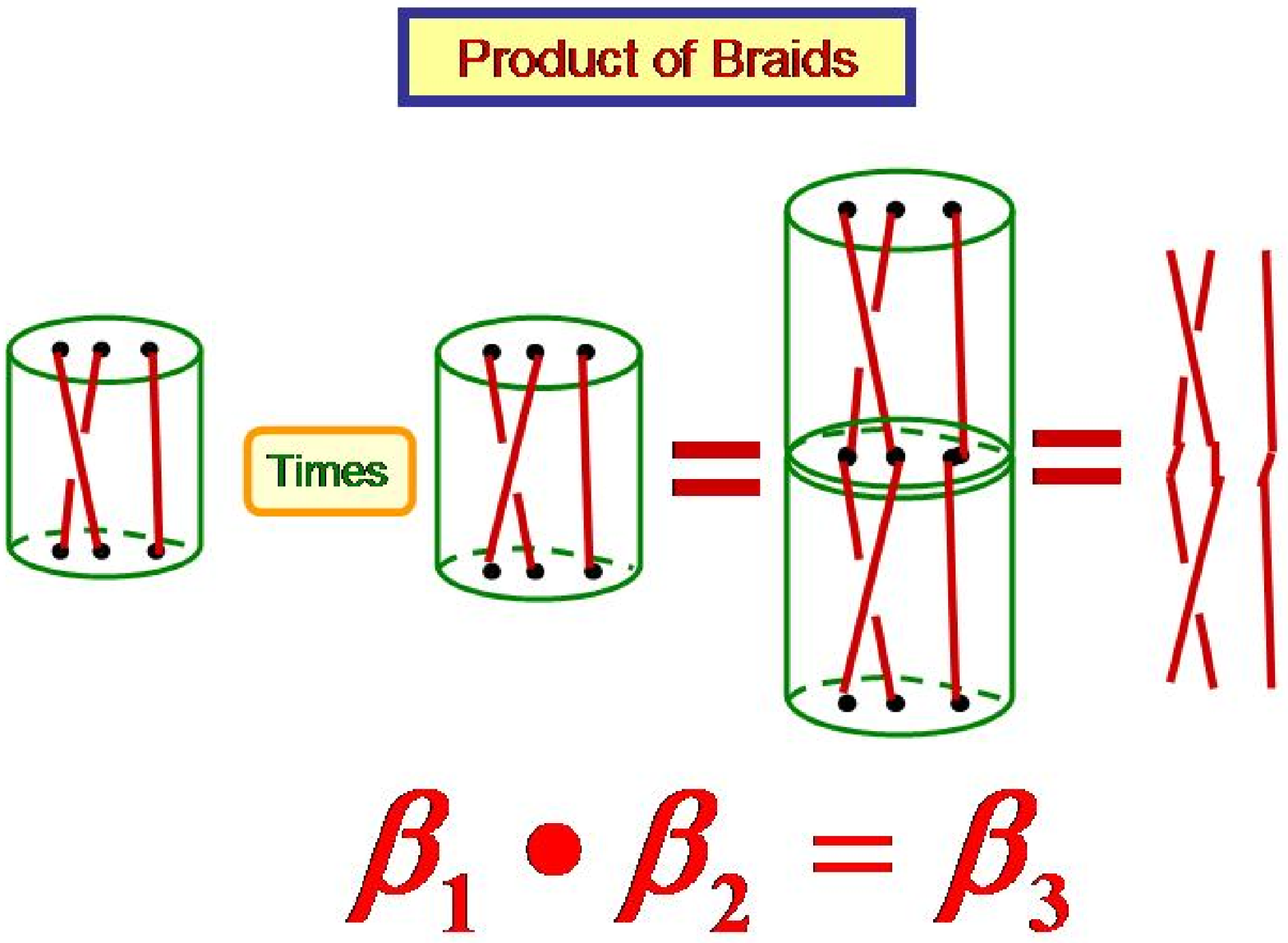}%
\\
\textbf{Figure 8. The product of braids.}%
\end{center}}}%
\qquad\qquad%
{\parbox[b]{2.5278in}{\begin{center}
\includegraphics[
natheight=7.499600in,
natwidth=9.999800in,
height=1.9026in,
width=2.5278in
]%
{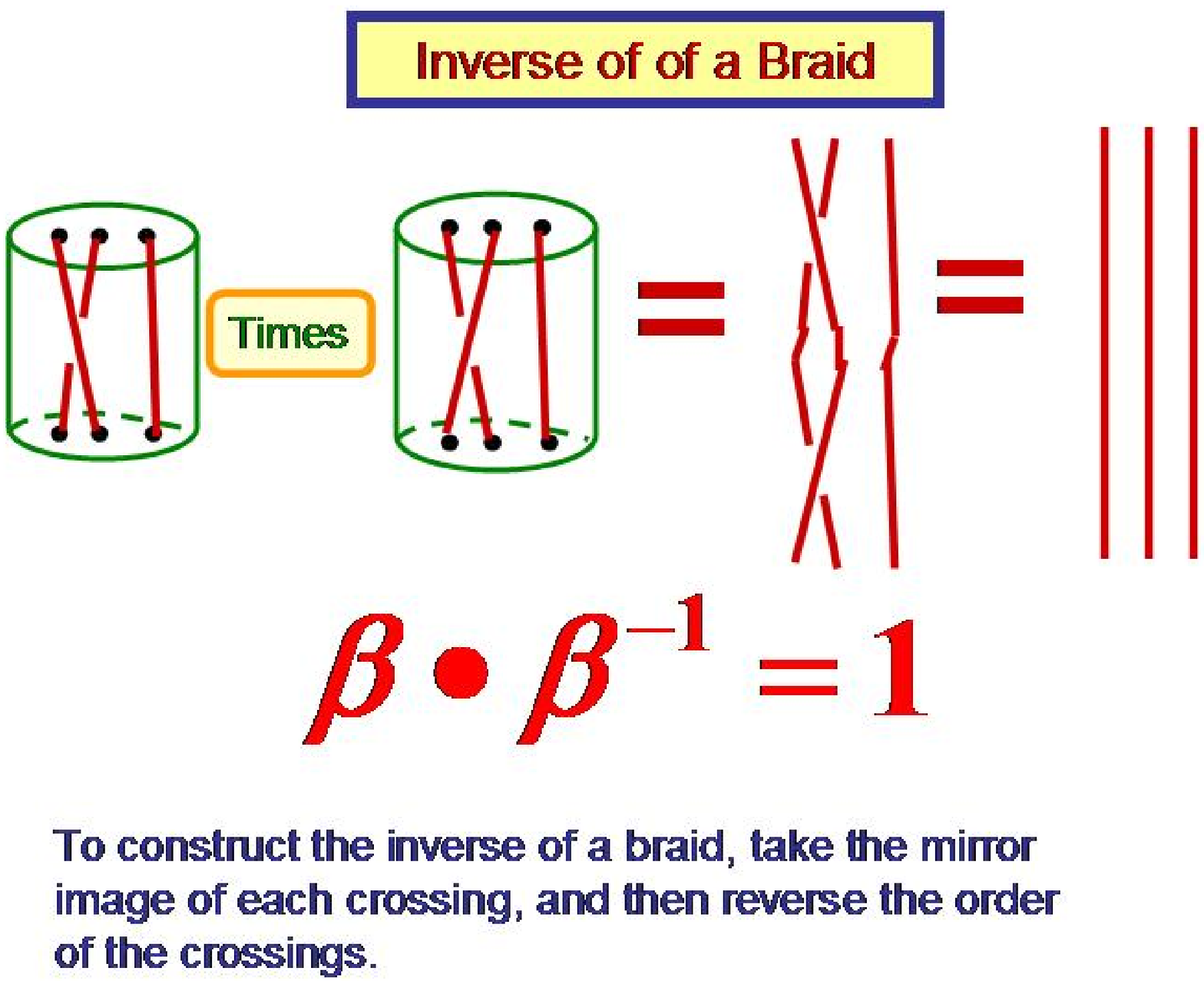}%
\\
\textbf{Figure 9. The inverse of a braid.}%
\end{center}}}%

\qquad%
{\parbox[b]{2.5278in}{\begin{center}
\includegraphics[
natheight=7.499600in,
natwidth=9.999800in,
height=1.9026in,
width=2.5278in
]%
{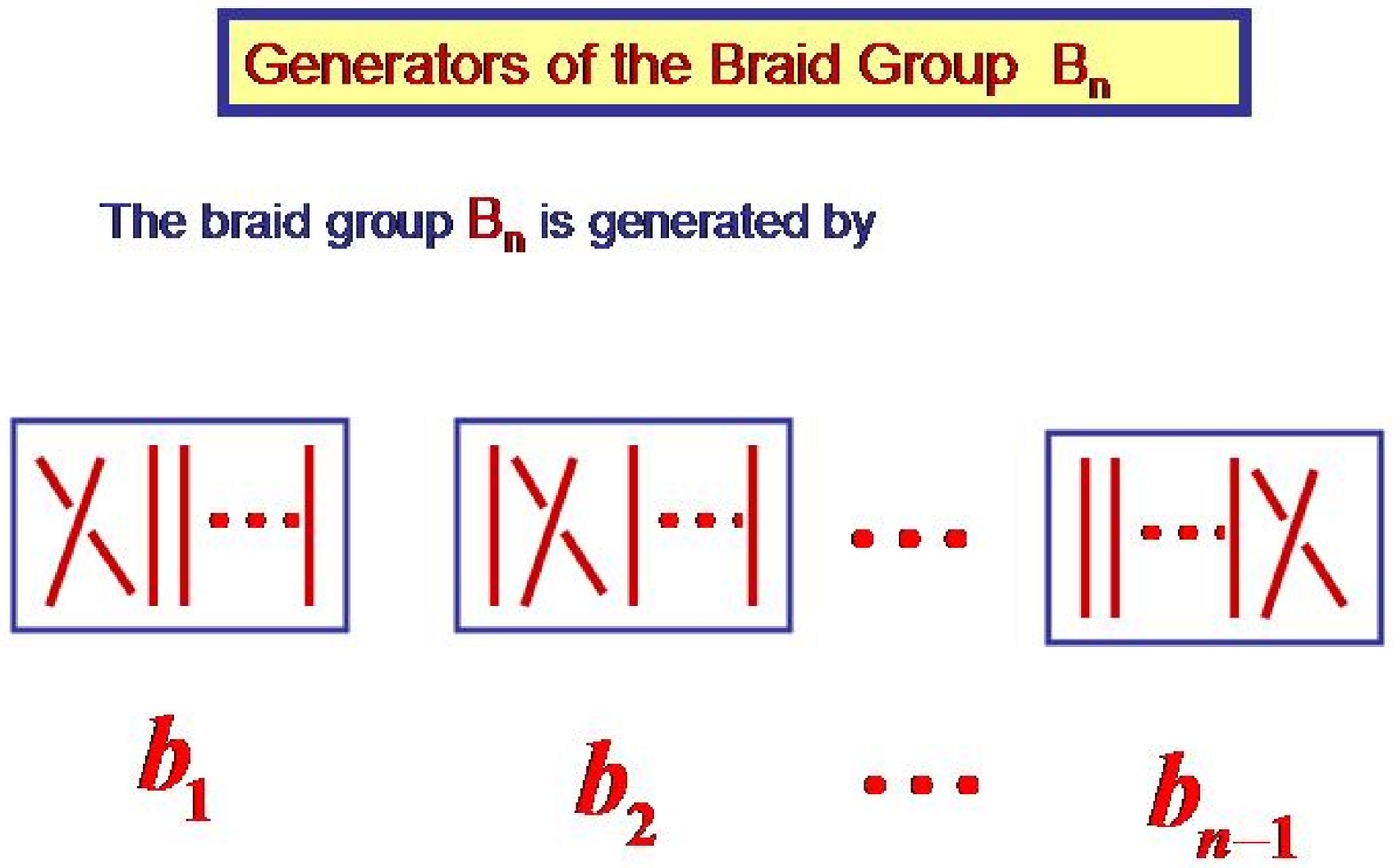}%
\\
\textbf{Figure 10. The generators of the braid group.}%
\end{center}}}%
\qquad\qquad%
{\parbox[b]{2.5278in}{\begin{center}
\includegraphics[
natheight=7.499600in,
natwidth=9.999800in,
height=1.9026in,
width=2.5278in
]%
{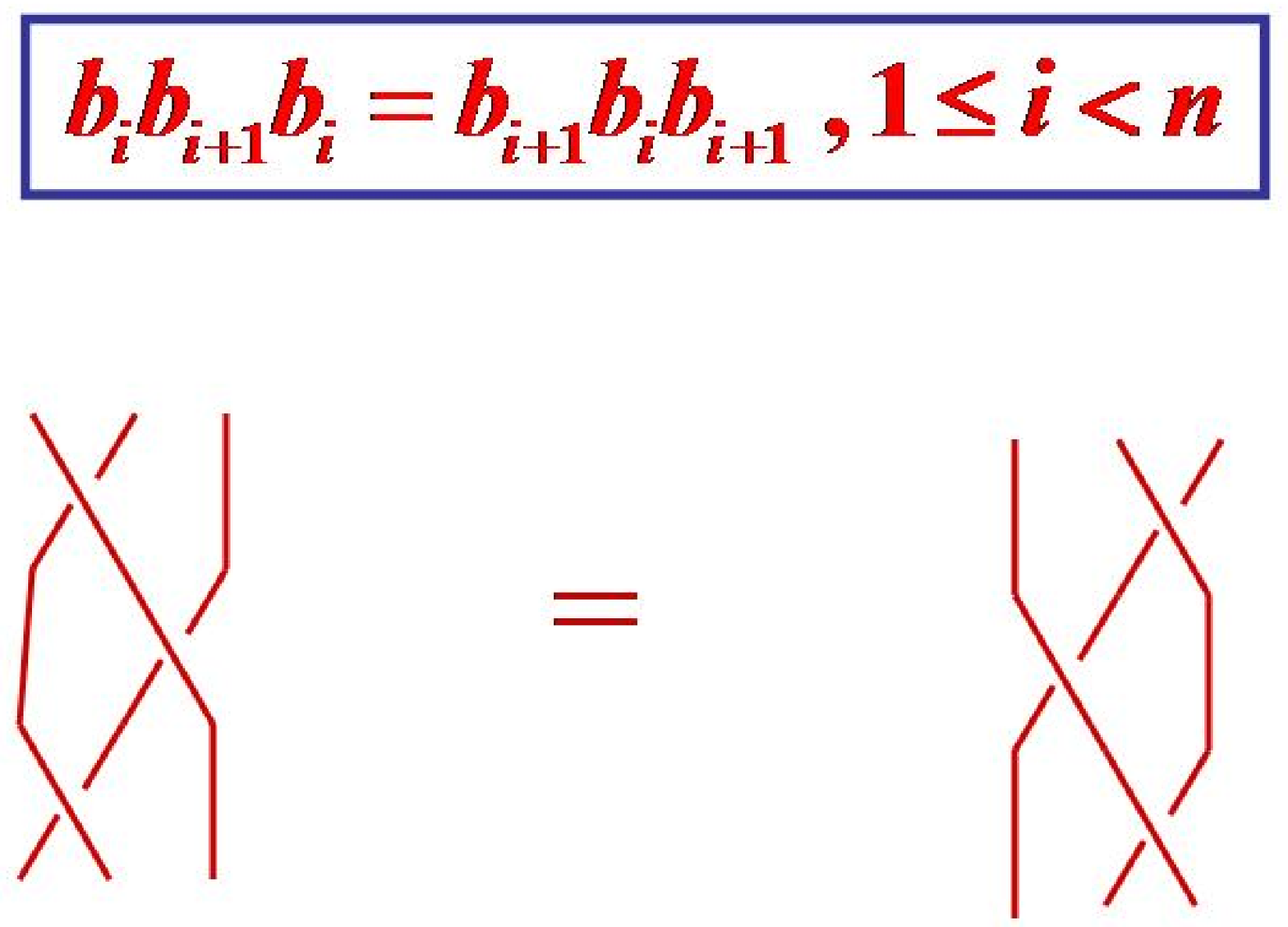}%
\\
\textbf{Figure 11. \ A diagrammatic illustration of one of the first set of
defining relations.}%
\end{center}}}%

\qquad%
{\parbox[b]{2.5278in}{\begin{center}
\includegraphics[
natheight=7.499600in,
natwidth=9.999800in,
height=1.9026in,
width=2.5278in
]%
{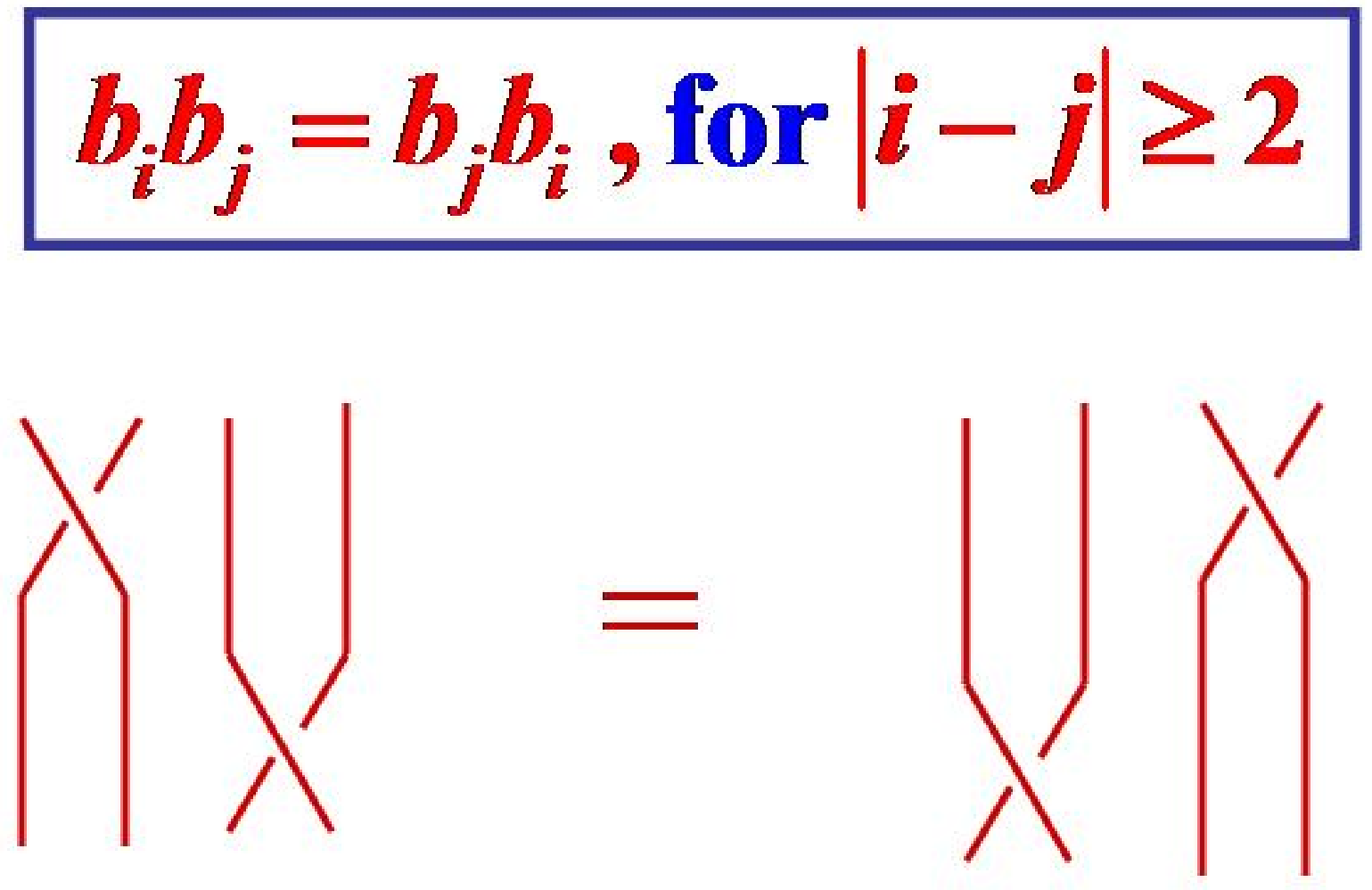}%
\\
\textbf{Figure 12. \ A diagrammatic illustration of one of the second set of
defining relations.}%
\end{center}}}%
\qquad\qquad%
{\parbox[b]{2.5278in}{\begin{center}
\includegraphics[
natheight=7.499600in,
natwidth=9.999800in,
height=1.9026in,
width=2.5278in
]%
{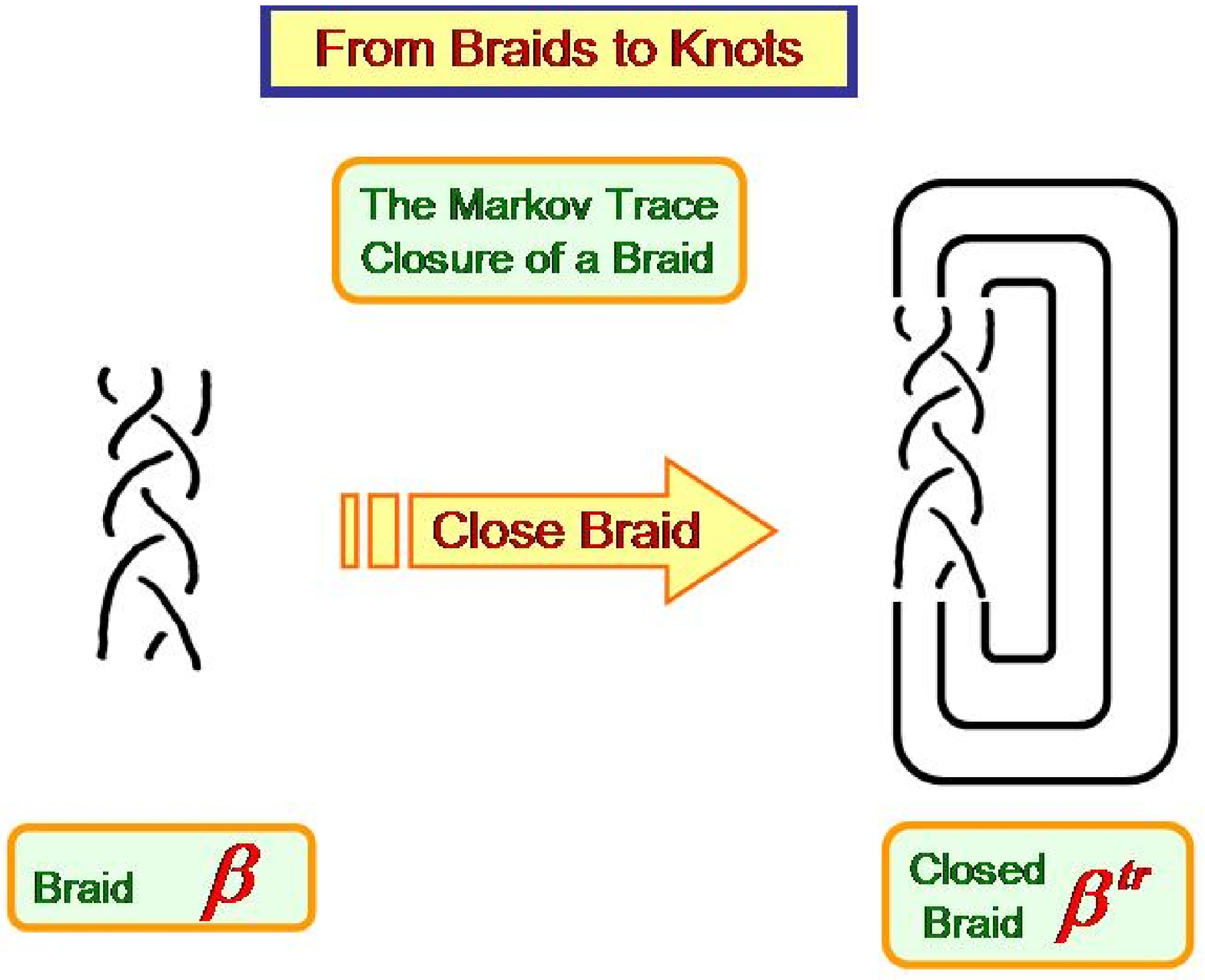}%
\\
\textbf{Figure 13. The closure (or trace) of a braid.}%
\end{center}}}%

\bigskip

\section{Transforming braids into knots and vice versa: The closure (or trace)
of a braid}

\bigskip

Why are braids of importance in knot theory?

The answer to this questions begins by observing that every braid $\beta$ can
be converted into a knot (or link) by forming the closure (a.k.a., trace)
$\beta^{tr}$ as shown in Figure 13. \ Then, of course, there is the famous
Markov theorem telling us when two braids produce the same knot (or link):

\bigskip

\begin{theorem}
[Markov]Two braids $\beta_{1}$ and $\beta_{2}$ produce the same knot (or link)
under braid closure if and only if there exists a finite sequence of Markov
moves that transforms one braid into the other.
\end{theorem}

\bigskip

The two Markov moves $M_{1}$ and $M_{2}$ are are shown in Figures 14 and 15.

\qquad%
{\parbox[b]{2.5278in}{\begin{center}
\includegraphics[
natheight=7.499600in,
natwidth=9.999800in,
height=1.9026in,
width=2.5278in
]%
{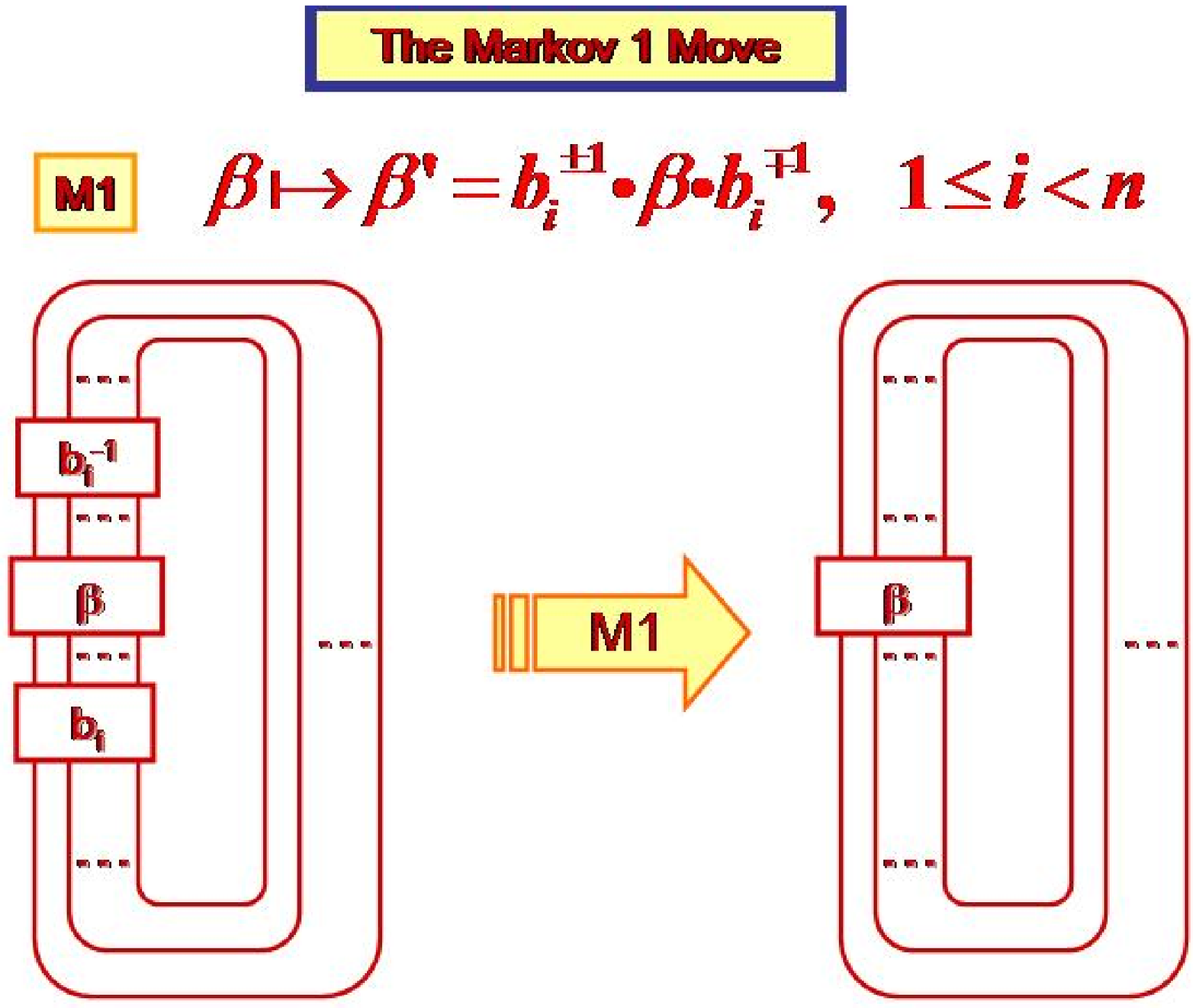}%
\\
\textbf{Figure 14. \ The Markov 1 move.}%
\end{center}}}%
\qquad\qquad%
{\parbox[b]{2.5278in}{\begin{center}
\includegraphics[
natheight=7.499600in,
natwidth=9.999800in,
height=1.9026in,
width=2.5278in
]%
{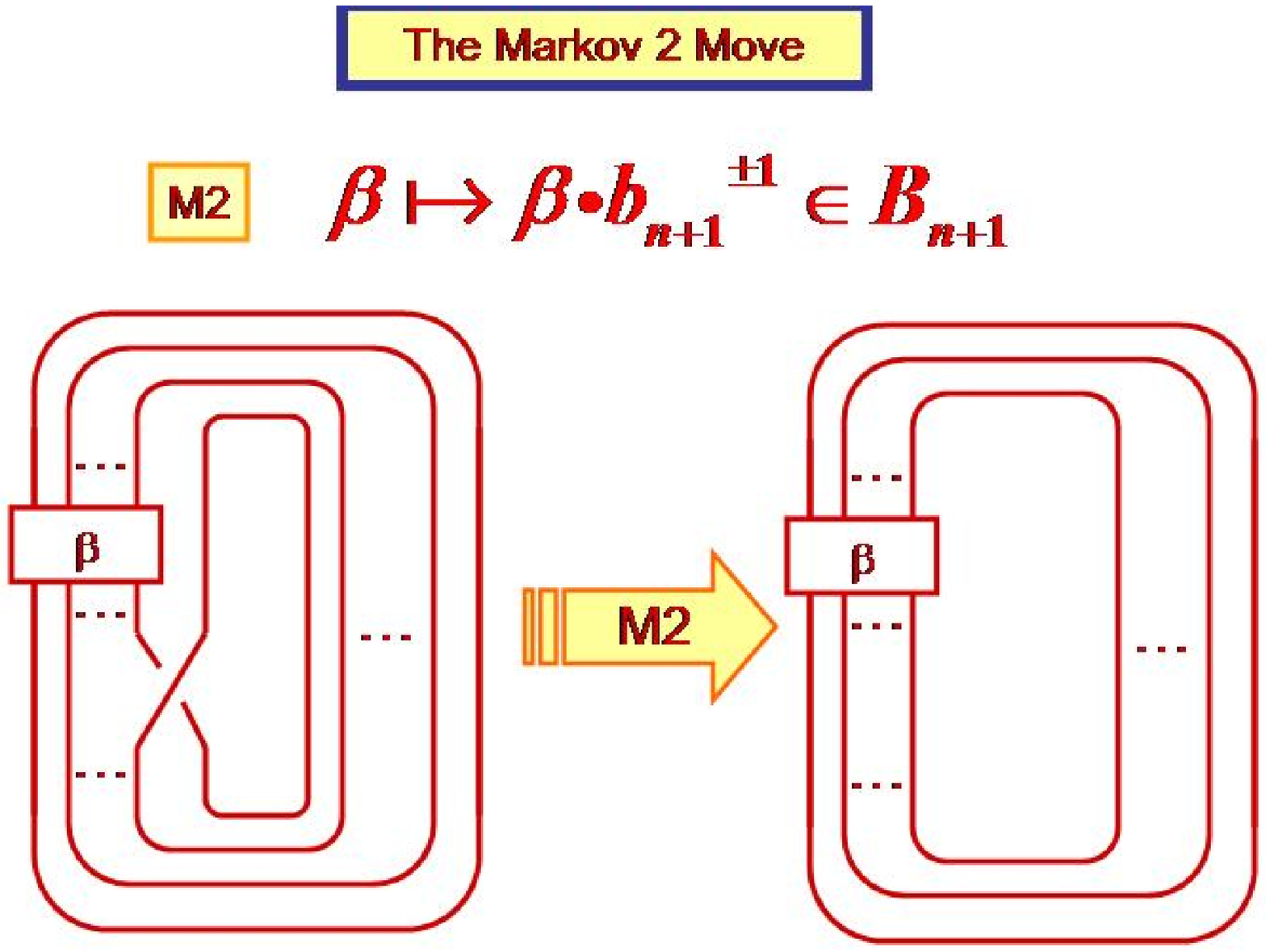}%
\\
\textbf{Figure 15. \ The Markov 2 move.}%
\end{center}}}%

\bigskip

Most amazingly, the process of transforming a braid into a knot (or link) can
be reversed, as stated by Alexander's theorem:

\bigskip

\begin{theorem}
[Alexander]Every knot (or link) is the closure of a braid.
\end{theorem}

\bigskip

Returning to our original highly algebraic definition of the braid group
$B_{n}$, we should mention that each braid $\beta$ in $B_{n}$ can be expressed
as a product of the generators $b_{1},b_{2},\ldots,b_{n-1}$ and their inverses
$b_{1}^{-1},b_{2}^{-1},\ldots,b_{n-1}^{-1}$. \ Thus, each braid $\beta$ can be
written in the form%
\[
\beta=%
{\displaystyle\prod\limits_{i=1}^{\ell}}
b_{j\left(  i\right)  }^{\epsilon\left(  i\right)  }=b_{j(1)}^{\epsilon
(1)}b_{j(2)}^{\epsilon(2)}\cdots b_{j(\ell)}^{\epsilon(\ell)}\text{ \ ,}%
\]
where $\epsilon\left(  i\right)  =\pm1$ for $i=1,2,\ldots,\ell$. \ We will
call such a product in the generators $b_{1},b_{2},\ldots,b_{n-1}$ and their
inverses $b_{1}^{-1},b_{2}^{-1},\ldots,b_{n-1}^{-1}$ a \textbf{word} defining
the braid $\beta$. \ Two words define the same braid $\beta$ if and only if it
is possible transform one into the other by applying a finite sequence of
Tieze transformations. \ For a definition of Tieze transformations, we refer
the reader to Crowell and Fox\cite{Crowell1}.

\bigskip

Finally, we define the \textbf{writhe} of a braid $\beta$, written
$Writhe\left(  \beta\right)  $, as the sum of the exponents in any word
defining the braid.

\bigskip

\section{The Temperley-Lieb algebra $TL_{n}\left(  d\right)  $}

\bigskip

Our next stepping stone to the definition of the Jones polynomial is an
algebra, called the Temperley-Lieb algebra.

\bigskip

\begin{definition}
Let $d$ be an indeterminate complex number. \ Then for each positive integer
$n$, the \textbf{Temperley-Lieb algebra} $TL_{n}\left(  d\right)  $ is defined
as the algebra with identity $1$ generated by the identity $1$ and the
symbols
\[
E_{1},E_{2},\ldots,E_{n-1}%
\]
subject to the defining relations%
\[
\left\{
\begin{array}
[c]{ll}%
E_{i}E_{j}=E_{j}E_{i} & \text{for \ }\left\vert i-j\right\vert \geq2\\
& \\
E_{i}E_{i\pm1}E_{i}=E_{i} & \text{for \ }1\leq i<n\\
& \\
E_{i}^{2}=dE_{i} & \text{for \ }1\leq i<n
\end{array}
\right.
\]

\end{definition}

The Temperley-Lieb algebra $TL_{n}\left(  d\right)  $ is easily understood in
terms of diagrammatics\footnote{The diagrammatic representation of the
Temperler-Leib algebra is due to Kauffman.\cite{Kauffman1}$^{,}$%
\cite{Kauffman3}$^{,}$\cite{Kauffman4}}, as illustrated in Figures 4 through
13. \ 

\bigskip

As shown in Figure 13, an element of the Temperley-Lieb algebra $TL_{3}\left(
d\right)  $ can be thought of as the algebra consisting of all linear
combinations of rectangles, each rectangle with $3$ points at the top and $3$
points at the bottom connected by smooth non-intersecting curves, called
\textbf{strands}. \ As shown in Figures 17 and 18, two such elements of
$TL_{n}\left(  d\right)  $ are equal if it is possible to continuously move
the strands of one into the strands of the other without cutting or breaking
the strands and without letting the strands pass through each other. \ Figure
17 shows two equal elements, and Figure 18 shows two non-equal elements. \ As
shown in Figure 19, the enclosing rectangle is usually omitted, but understood
to be there.

\bigskip

The product of two elements of $TL_{n}\left(  d\right)  $ is defined, as shown
in Figure 20, i.e., by stacking one rectangle on top of the other. \ As
illustrated in Figure 21, if a circle should happen to arise as a result of
the product, then it is simply replaced by the indeterminate $d$ times the
resulting rectangle with the circle omitted. \ The generators of the
Temperley-Lieb algebra are shown in Figure 22. \ We leave, as an amusing
exercise for the reader, the task of translating the complete set of defining
relations for the Temperley-Leib algebra given above into diagrammatics. \ 

\bigskip

\qquad%
{\parbox[b]{2.5278in}{\begin{center}
\includegraphics[
natheight=7.499600in,
natwidth=9.999800in,
height=1.9026in,
width=2.5278in
]%
{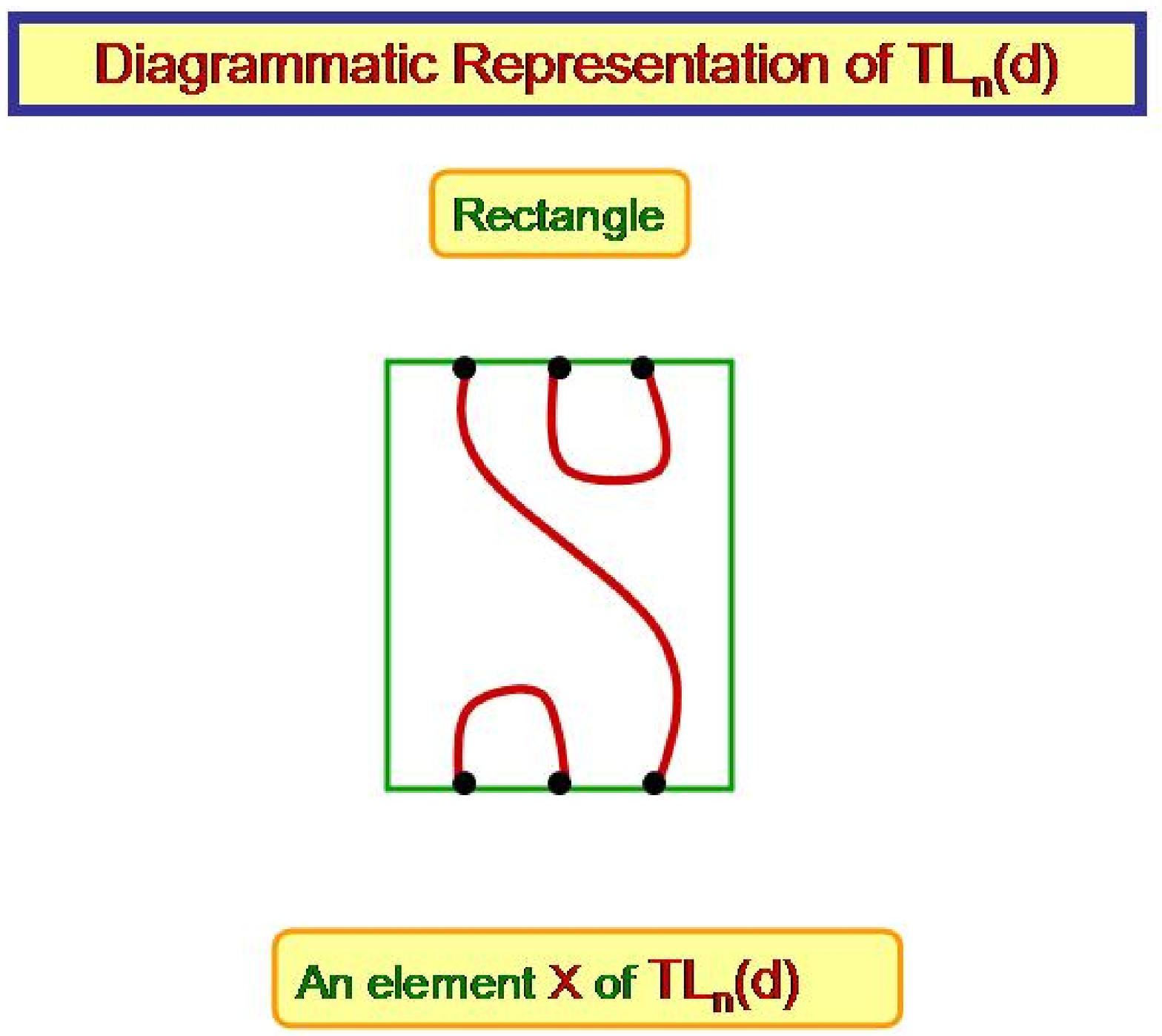}%
\\
\textbf{Figure 16.}%
\end{center}}}%
\qquad\qquad%
{\parbox[b]{2.5278in}{\begin{center}
\includegraphics[
natheight=7.499600in,
natwidth=9.999800in,
height=1.9026in,
width=2.5278in
]%
{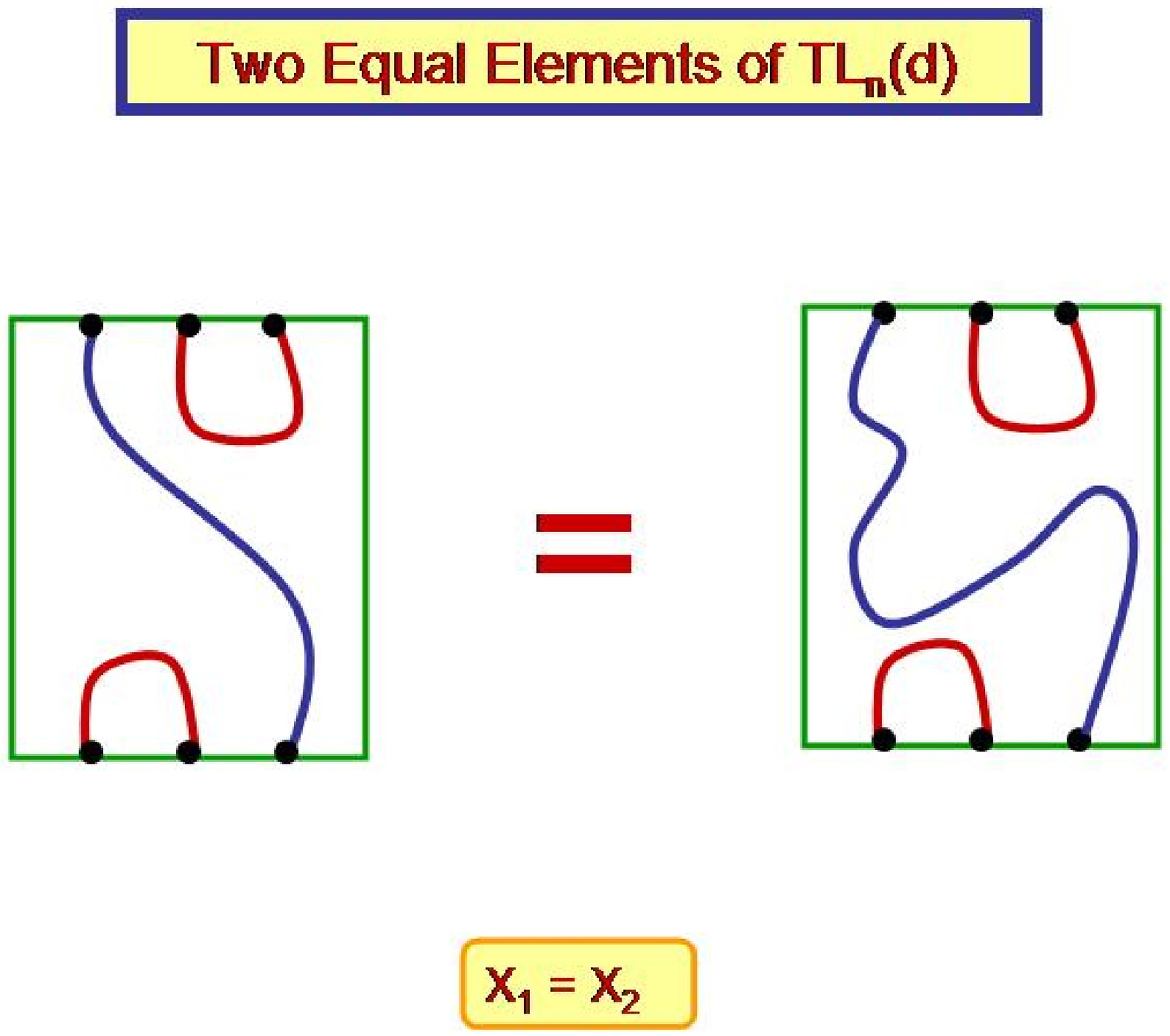}%
\\
\textbf{Figure 17.}%
\end{center}}}%

\qquad%
{\parbox[b]{2.5278in}{\begin{center}
\includegraphics[
natheight=7.499600in,
natwidth=9.999800in,
height=1.9026in,
width=2.5278in
]%
{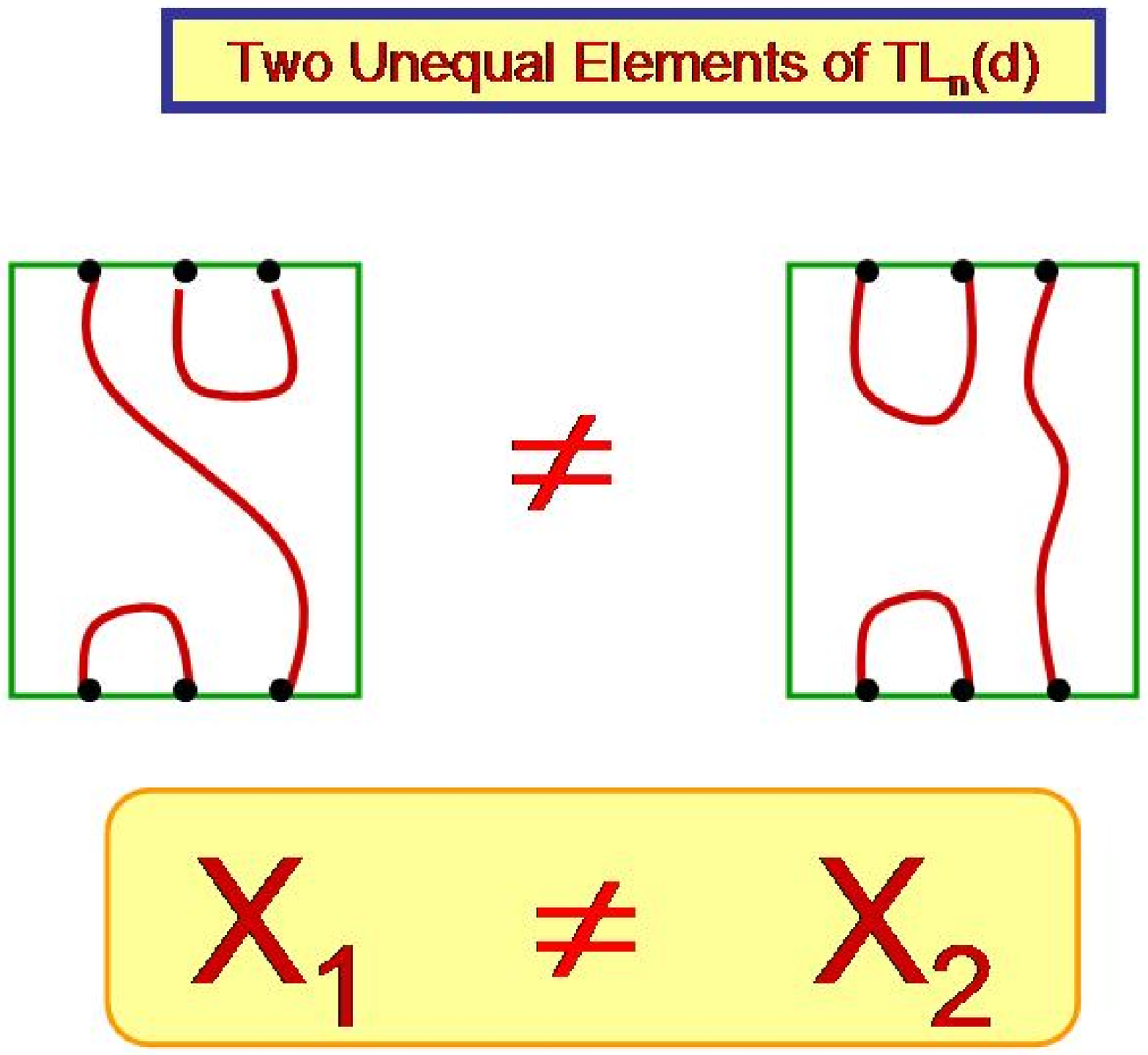}%
\\
\textbf{Figure 18.}%
\end{center}}}%
\qquad\qquad%
{\parbox[b]{2.5278in}{\begin{center}
\includegraphics[
natheight=7.499600in,
natwidth=9.999800in,
height=1.9026in,
width=2.5278in
]%
{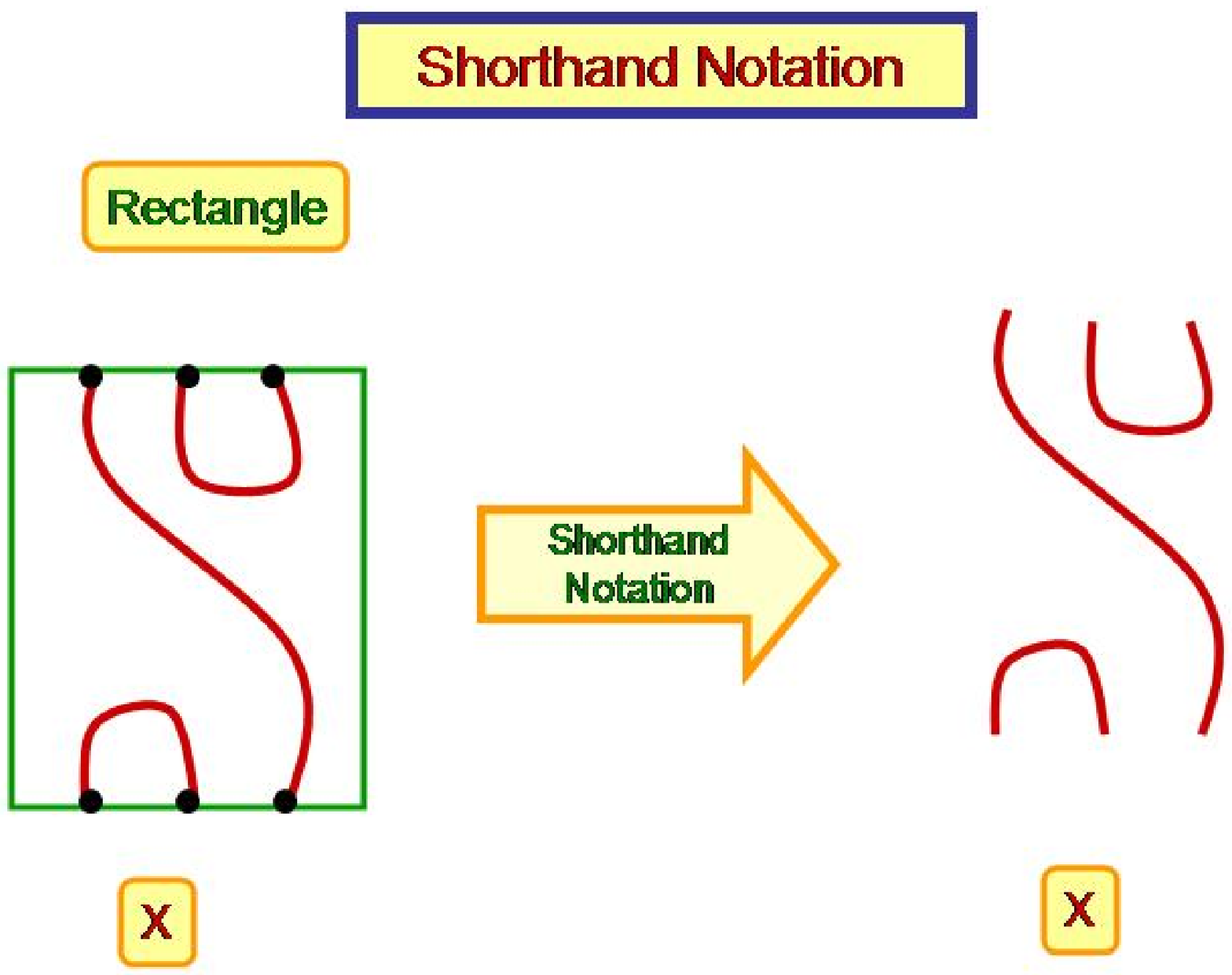}%
\\
\textbf{Figure 19.}%
\end{center}}}%

\qquad%
{\parbox[b]{2.5278in}{\begin{center}
\includegraphics[
natheight=7.499600in,
natwidth=9.999800in,
height=1.9026in,
width=2.5278in
]%
{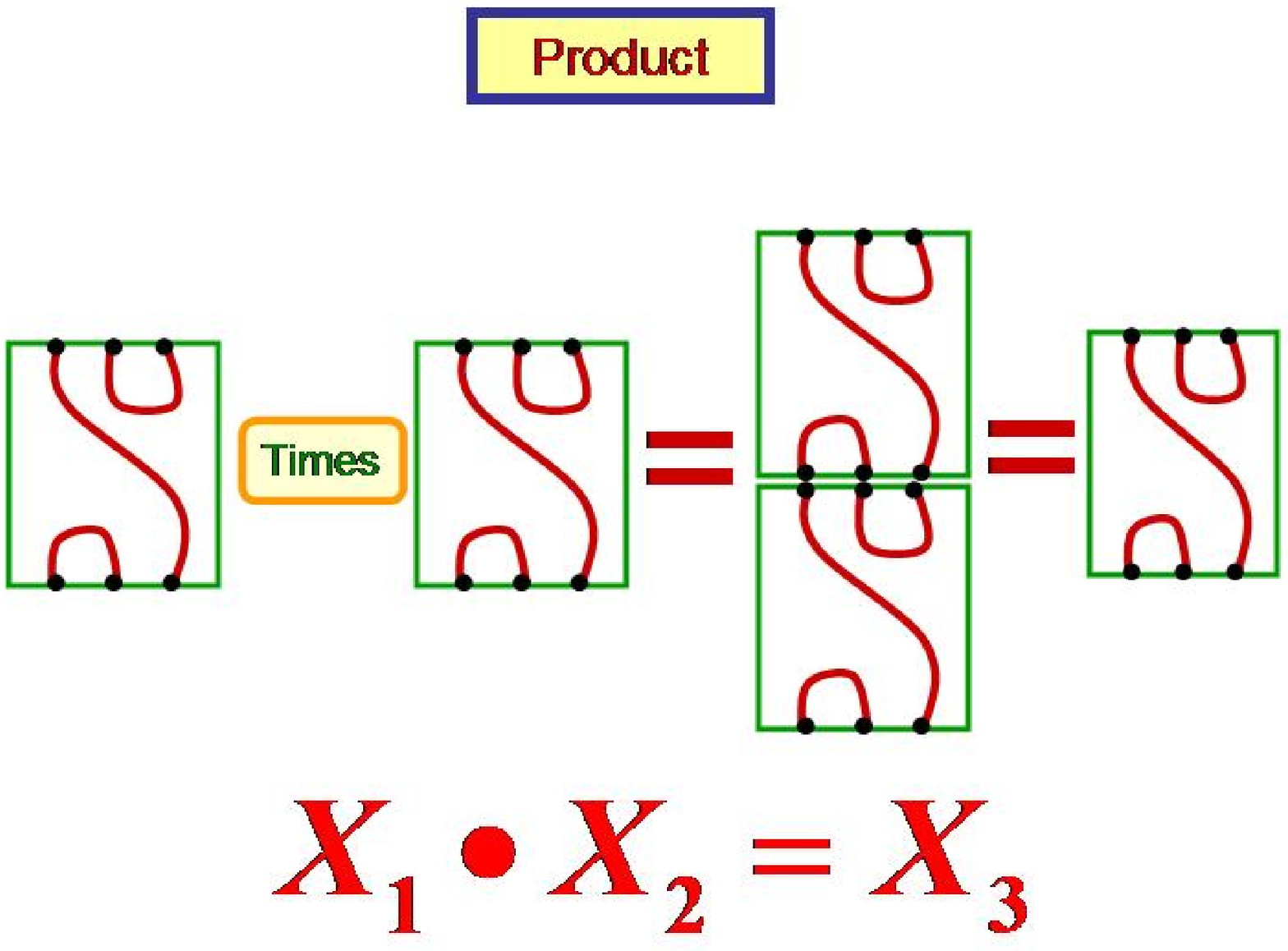}%
\\
\textbf{Figure 20.}%
\end{center}}}%
\qquad\qquad%
{\parbox[b]{2.5278in}{\begin{center}
\includegraphics[
natheight=7.499600in,
natwidth=9.999800in,
height=1.9026in,
width=2.5278in
]%
{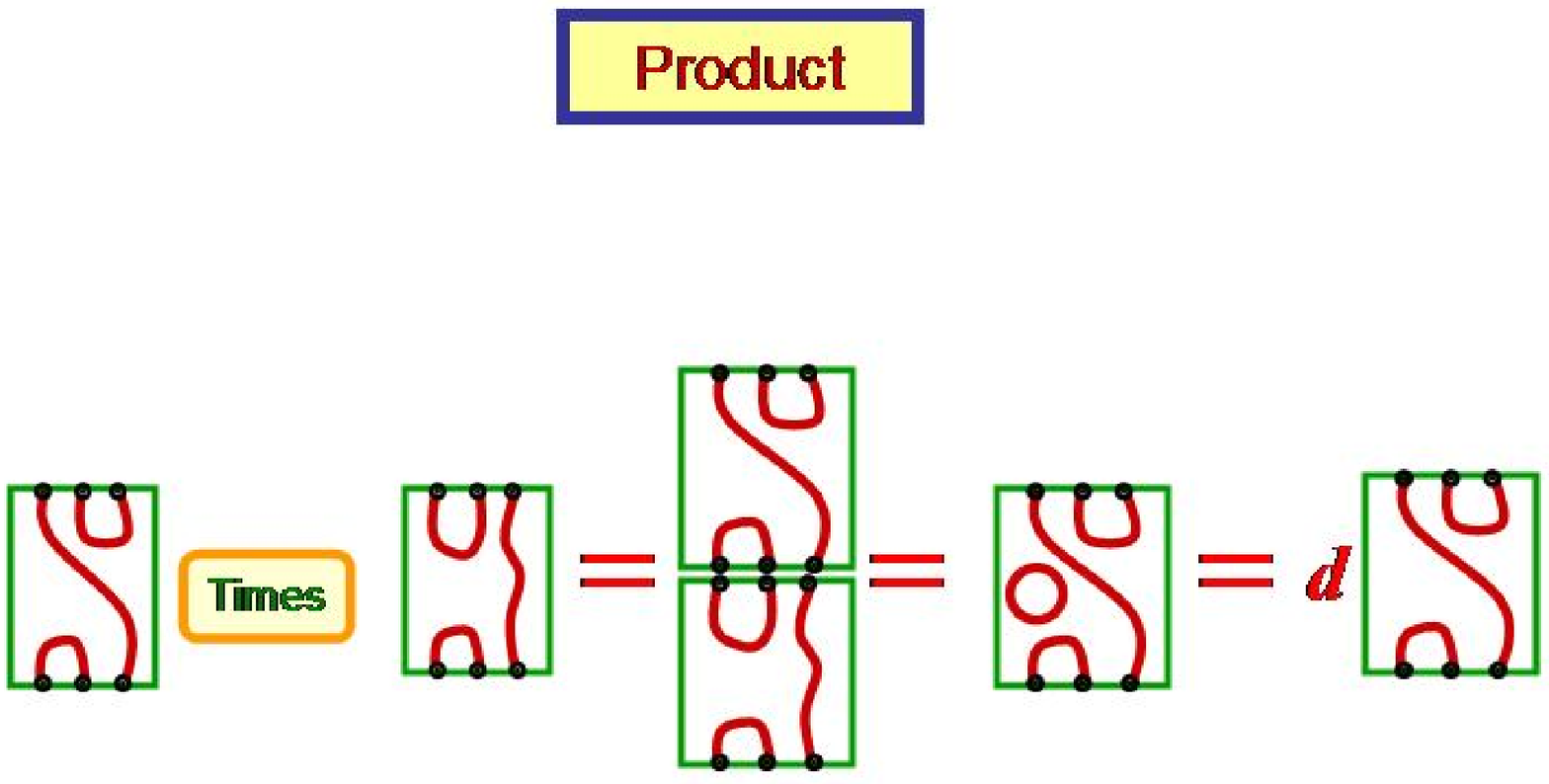}%
\\
\textbf{Figure 21.}%
\end{center}}}%

\qquad%
{\parbox[b]{2.5278in}{\begin{center}
\includegraphics[
natheight=7.499600in,
natwidth=9.999800in,
height=1.9026in,
width=2.5278in
]%
{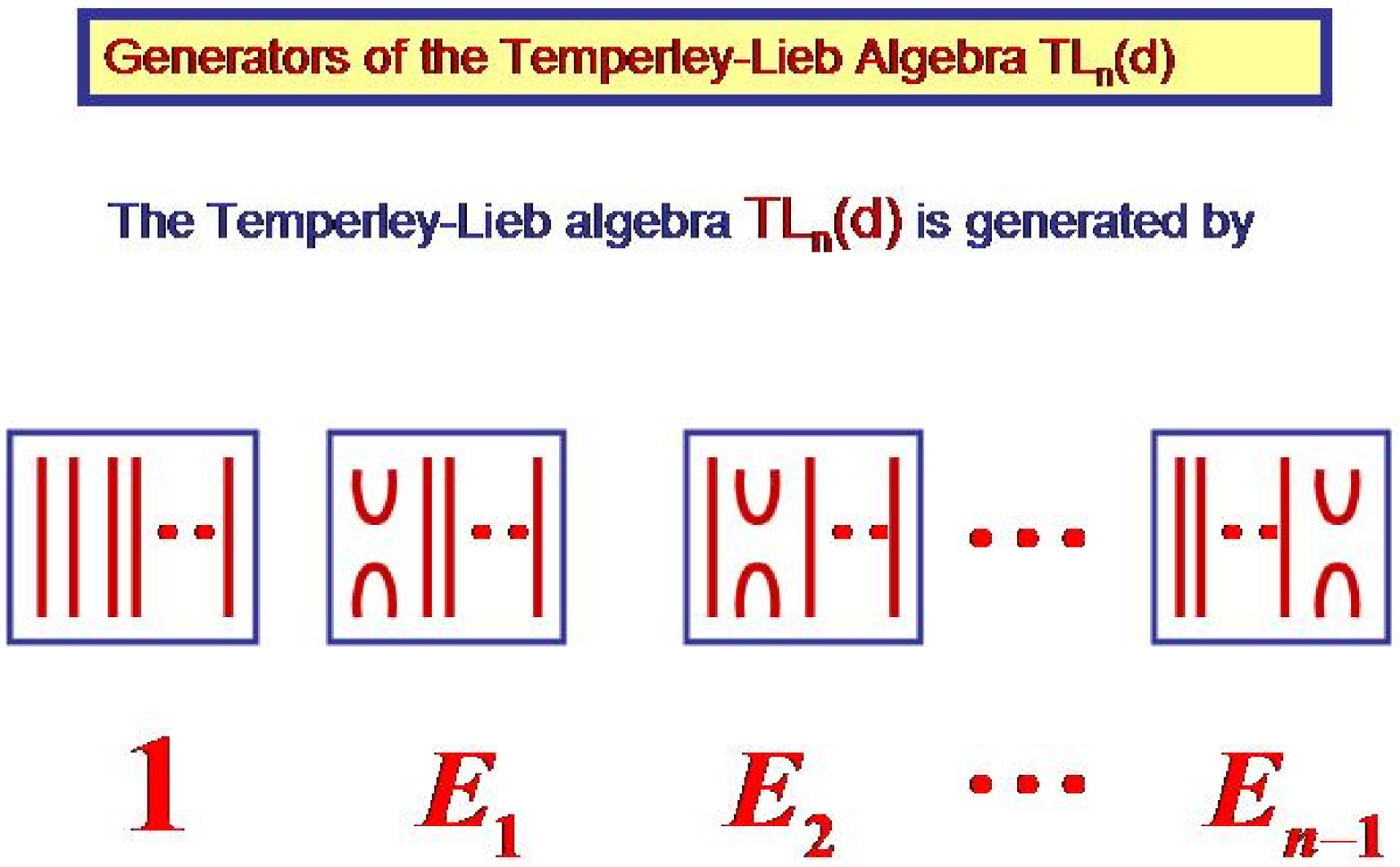}%
\\
\textbf{Figure 22. \ The Temperley-Lieb algebra }$\mathbf{TL}_{n}\left(
d\right)  $\textbf{ is generated by all formal sums of products of the above
generators.}%
\end{center}}}%
\qquad\qquad%
{\parbox[b]{2.5278in}{\begin{center}
\includegraphics[
natheight=7.499600in,
natwidth=9.999800in,
height=1.9026in,
width=2.5278in
]%
{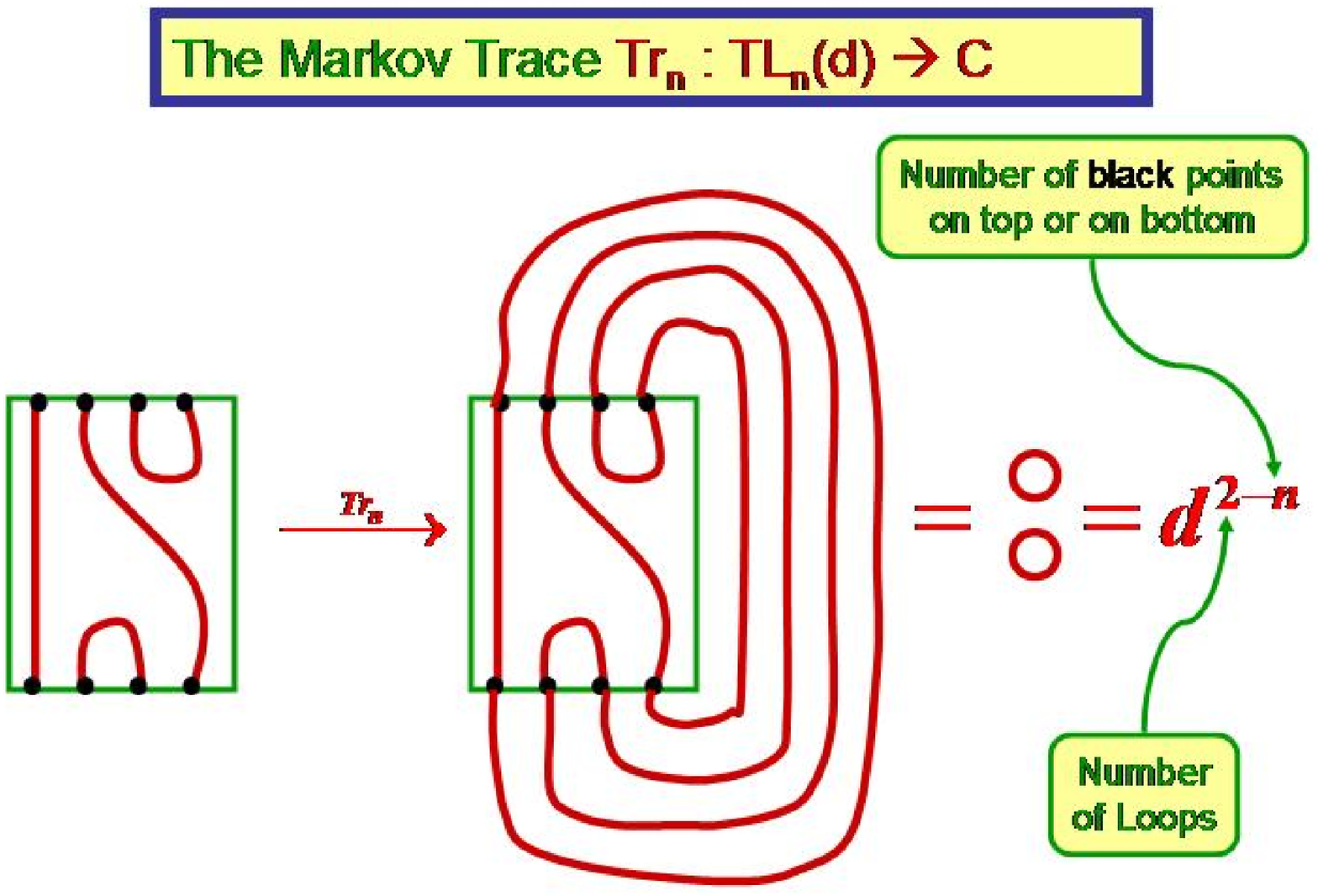}%
\\
\textbf{Figure 23. The diagramatic definition of the Markov trace }%
$Tr_{n}:TL_{n}\left(  d\right)  \longrightarrow\mathbb{C}$.
\end{center}}}%

We would be amiss if we did not mention that there is a map $Tr_{n}%
:TL_{n}\left(  d\right)  \longrightarrow\mathbb{C}$ from the Temperley-Lieb
algebra $TL_{n}\left(  d\right)  $ to the complex numbers $\mathbb{C}$, called
the \textbf{Markov trace}, satisfying the following three conditions:

\begin{itemize}
\item $Tr_{n}\left(  1\right)  =1$

\item $Tr_{n}\left(  XY\right)  =Tr_{n}\left(  YX\right)  $ for all $X$ and
$Y$ in $TL_{n}\left(  d\right)  $

\item If $X\in TL_{n}\left(  d\right)  $, then $Tr_{n+1}\left(  XE_{n}\right)
=\frac{1}{d}Tr_{n}\left(  X\right)  $
\end{itemize}

\bigskip

A diagrammatic definition of the Markov trace is shown in figure 23.

\bigskip

We will later need the following theorem:

\bigskip

\begin{theorem}
The above three conditions uniquely determine the Markov trace, i.e., any map
$TL_{n}\left(  d\right)  \longrightarrow\mathbb{C}$ satisfying the above three
conditions must be the same as that defined by figure 23.
\end{theorem}

\bigskip

\section{The definition of the Jones polynomial}

\bigskip

Let $d$ be an indeterminate complex number, and let $A$ also be an
indeterminant complex number such that $d=-A^{2}-A^{-2}$. Let $TL_{n}\left(
d\right)  $ be the corresponding Temperley-Lieb algebra, and let $B_{n}$
denote the $n$-stranded braid group. \ Then the \textbf{Jones representation}
\[
\rho_{A}:TL_{n}\left(  d\right)  \longrightarrow B_{n}%
\]
is the group representation defined by
\[
\left\{
\begin{array}
[c]{lll}%
b_{i} & \longmapsto & AE_{i}+A^{-1}1\\
&  & \\
b_{i}^{-1} & \longmapsto & A^{-1}E_{i}+A1
\end{array}
\right.
\]
where $b_{1},b_{2},\ldots,b_{n-1}$ denote the generators of the braid group
$B_{n}$, and where $1,E_{1},E_{2},\ldots,E_{n-1}$ denote the generators of the
Temperley-Lieb algebra $TL_{n}\left(  d\right)  $. \ 

\bigskip

We leave for the reader's amusement the exercise of verifying that the images
under $\rho_{A}$ of the generators $b_{1},b_{2},\ldots,b_{n-1}$ satisfy the
defining relations of the braid group.

\bigskip

We are now finally in a position to define the Jones polynomial.

\bigskip

Let $\beta$ be an element of the $n$-stranded braid group $B_{n}$, and let
$\beta^{Tr}$ denote the knot (or link) constructed from the closure of the
braid $\beta$. \ Then the \textbf{Jones polynomial} $V_{\beta^{Tr}}\left(
t\right)  $ of the knot (or link) $\beta^{Tr}$ is the Laurent polynomial in
the polynomial ring $\mathbb{Z}\left[  t,t^{-1}\right]  $ over the integers
$\mathbb{Z}$ given by
\[
V_{\beta^{Tr}}\left(  A^{-4}\right)  =-A^{2Writhe\left(  \beta\right)
}d^{n-1}Tr_{n}\left(  \rho_{A}\left(  \beta\right)  \right)  \text{ \ ,}%
\]
where $t=A^{-4}$, where $Writhe\left(  \beta^{Tr}\right)  $ denotes the
writhe\footnote{Please refer to section 2 for a definition of writhe.} of the
braid $\beta$, and where $Tr_{n}\left(  \rho_{A}\left(  \beta\right)  \right)
$ denotes the Markov trace of the value of the Jones representation $\rho_{A}$
on the braid $\beta$.

\bigskip

\section{The representation $\Phi:TL_{n}\left(  d\right)  \longrightarrow
\mathbb{C}U\left(  \mathcal{H}_{n,k}\right)  $ of the Temperley-Lieb algebra
$TL_{n}\left(  d\right)  $}

\bigskip

Our objective is to describe the polytime quantum algorithm in
AJL\cite{Aharonov1} for approximating values of the Jones polynomial
$V_{\beta^{Tr}}\left(  t\right)  $ at the primitive $k$-th roots of unity
$t=e^{2\pi i/k}$ for positive integers $k$. \ To this end, we begin by
constructing a representation of the Temperley-Lieb algebra $TL_{n}\left(
d\right)  $ which carries the image of the Jones representation $\rho
_{A}:B_{n}\longrightarrow TL_{n}\left(  d\right)  $ onto a group of unitary transformations.

\bigskip

Let $G_{k}$ denote the graph of $k-1$ vertices and $k-2$ edges given in Figure 24.%

\begin{center}
\includegraphics[
trim=0.000000in 4.415014in 4.215915in 0.000000in,
natheight=7.499600in,
natwidth=9.999800in,
height=0.7982in,
width=1.4736in
]%
{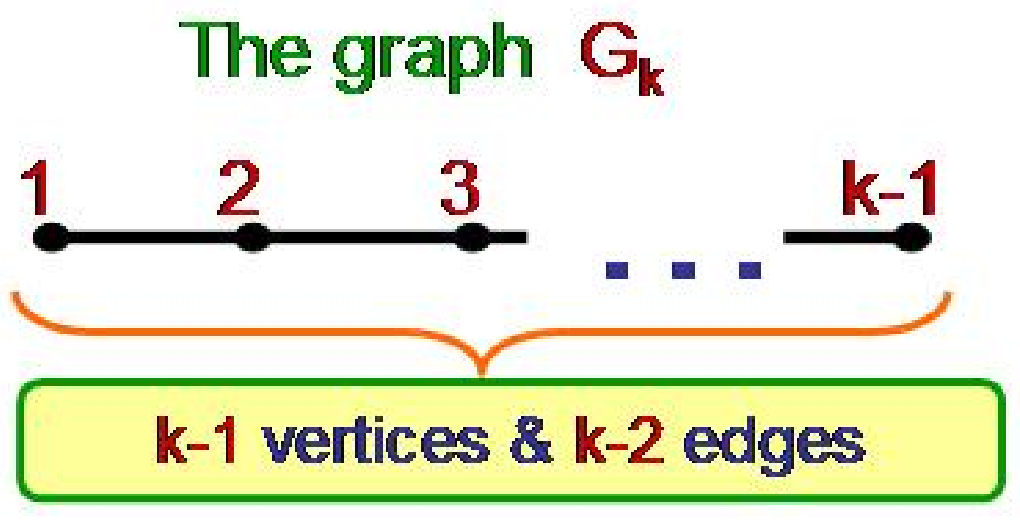}%
\\
\textbf{Figure 24. \ The graph }$G_{k}$\textbf{ of }$k-1$\textbf{ vertices and
}$k-2$\textbf{ edges.}%
\end{center}

\bigskip

The adjacency matrix $M_{k}$ of the graph $G_{k}$ is easily seen to be
\[
M_{k}=\left(
\begin{array}
[c]{rrrrrr}%
0 & 1 & 0 & \cdots & 0 & 0\\
1 & 0 & 1 &  & 0 & 0\\
0 & 1 & 0 &  &  & \\
& \vdots &  & \ddots &  & \\
0 & 0 & 0 &  & 0 & 1\\
0 & 0 & 0 & \cdots & 1 & 0
\end{array}
\right)
\]
Moreover, $d=2\cos\left(  \pi/k\right)  $ can be shown to be an eigenvalue of
$M_{k}$ corresponding to the eigenvector
\[
\overrightarrow{\lambda}=\left(  \lambda_{1},\lambda_{2},\ldots,\lambda
_{k-1}\right)  =\left(  \ \sin\left(  \pi\ell/k\right)  \ \right)  _{1\leq
\ell<k}%
\]

\bigskip

We can now construct our representation as follows:

\bigskip

Let\text{ }$\mathcal{P}_{n,k}$ denote the set of paths in the graph $G_{k}$ of
\textbf{length} $n$ starting at the vertex $1$, and let $\mathcal{H}_{n,k}$
denote the Hilbert space with orthonormal basis
\[
\left\{  \ \left\vert p\right\rangle :p\in\mathcal{P}_{n,k}\right\}
\]
with basis elements labelled by the paths $p$ in $\mathcal{P}_{n,k}$.

\bigskip

Interpreting $0$ as "to the left" and $1$ as "to the right," we identify each
length $n$ path $p$ with a binary string of length $n$. \ 

\bigskip

For each path $p$, let $p^{\left.  i-1\right\rfloor }$ be the subpath
corresponding to the first $i-1$ bits of $p$, let $p^{\left\lfloor i\cdots
i+1\right\rfloor }$ denote the subpath corresponding to bits $i$ up to and
including bit $i+1$ of $p$, and finally let $p^{\left\lfloor i+2\right.  }$
denote the subpath of $p$ corresponding to bits $i+2$ up to and including the
last $n$-th bit. \ Let $e_{i}\left(  p\right)  $ be the endpoint of the
subpath $p^{\left.  i-1\right\rfloor }$. \ Hence, $e_{i}\left(  p\right)
\in\left\{  1,2,\ldots,k-1\right\}  $.

\bigskip

Select $d=2\cos\left(  \pi/k\right)  $. \ Since $A$ is related to $d$ via the
formula $d=-A^{2}-A^{-2}$, we choose among the first four possible choices
$\pm e^{i\pi/\left(  2k\right)  }$ for $A$ the value
\[
A=e^{-i\pi/\left(  2k\right)  }%
\]

\bigskip

We are now ready to define a representation%
\[
\Phi:TL_{n}\left(  d\right)  \longrightarrow\mathbb{C}U\left(  \mathcal{H}%
_{n,k}\right)
\]
of the Temperley-Lieb algebra $TL_{n}\left(  d\right)  $ into the group ring
$\mathbb{C}U\left(  \mathcal{H}_{n,k}\right)  $ of the group $U\left(
\mathcal{H}_{n,k}\right)  $ of unitary transformations on the Hilbert space
$\mathcal{H}_{n,k}$ by specifying the images
\[
\Phi_{i}=\Phi\left(  E_{i}\right)
\]
of each of the generators $E_{i}$ of the Temperley-Lieb algebra $TL_{n}\left(
d\right)  $, taking great care to make sure that the $\Phi_{i}$'s satisfy the
same defining relations as the $E_{i}$'s.

\bigskip

We define $\Phi_{i}$ as:
\[
\Phi_{i}\left\vert p\right\rangle =\left\{
\begin{array}
[c]{ll}%
0 & \text{if \ }p^{\left\lfloor i\cdots i+1\right\rfloor }=00\\
& \\
\frac{\lambda_{e_{i}\left(  p\right)  -1}}{\lambda_{e_{i}\left(  p\right)  }%
}\left\vert p\right\rangle +\frac{\sqrt{\lambda_{e_{i}\left(  p\right)
-1}\lambda_{e_{i}\left(  p\right)  +1}}}{\lambda_{e_{i}\left(  p\right)  }%
}\left\vert p^{\left.  i-1\right\rfloor }10p^{\left\lfloor i+2\right.
}\right\rangle  & \text{if \ }p^{\left\lfloor i\cdots i+1\right\rfloor }=01\\
& \\
\frac{\sqrt{\lambda_{e_{i}\left(  p\right)  -1}\lambda_{e_{i}\left(  p\right)
+1}}}{\lambda_{e_{i}\left(  p\right)  }}\left\vert p^{\left.  i-1\right\rfloor
}01p^{\left\lfloor i+2\right.  }\right\rangle +\frac{\lambda_{e_{i}\left(
p\right)  +1}}{\lambda_{e_{i}\left(  p\right)  }}\left\vert p\right\rangle  &
\text{if \ }p^{\left\lfloor i\cdots i+1\right\rfloor }=10\\
& \\
0 & \text{if \ }p^{\left\lfloor i\cdots i+1\right\rfloor }=11
\end{array}
\right.
\]
We leave for the reader the exercise of showing that the transformations
$\Phi_{i}$ satisfy the defining identities of the $E_{i}$'s, i.e., the
identities
\[
\left\{
\begin{array}
[c]{ll}%
\Phi_{i}\Phi_{j}=\Phi_{j}\Phi_{i} & \text{for \ }\left\vert i-j\right\vert
\geq2\\
& \\
\Phi_{i}\Phi_{i\pm1}\Phi_{i}=\Phi_{i} & \text{for \ }1\leq i<n\\
& \\
\Phi_{i}^{2}=d\Phi_{i} & \text{for \ }1\leq i<n
\end{array}
\right.
\]
and hence that $\Phi$ is a legitimate representation of the Temperley-Leib
algebra $TL_{n}\left(  d\right)  $.

\bigskip

\section{Constructing a trace $\widetilde{Tr}$ compatible with the Markov
trace $Tr_{n}$}

\bigskip

We next need to construct a trace $\widetilde{Tr}$ on the image of the
representation $\Phi:TL_{n}\left(  d\right)  \longrightarrow\mathbb{C}U\left(
\mathcal{H}_{n,k}\right)  $ which is compatible with the Markov trace $Tr_{n}%
$, i.e., a trace $\widetilde{Tr}$ such that the following diagram is
commutative%
\[%
\begin{array}
[c]{lcl}%
TL_{n}\left(  d\right)  & \quad\overset{\Phi}{\longrightarrow}\quad &
\operatorname{Im}\left(  \Phi\right)  \subset\mathbb{C}U\left(  \mathcal{H}%
_{n,k}\right) \\
& Tr_{n}\searrow\quad & \quad\downarrow\widetilde{Tr}\\
&  & \quad\ \mathbb{C}%
\end{array}
\]
For this construction, we need the following lemma:

\bigskip

\begin{lemma}
The representation $\Phi:TL_{n}\left(  d\right)  \longrightarrow
\mathbb{C}U\left(  \mathcal{H}_{n,k}\right)  $ maps each ket $\left\vert
p\right\rangle $ to a linear combination of kets each labeled by a path of the
same length as the path $p$, and each having the same endpoint as $p.$
\end{lemma}

\bigskip

An immediate corollary is:

\bigskip

\begin{corollary}
Let $\mathcal{P}_{n,k,m}$ be the subset of $\mathcal{P}_{n,k}$ of all paths
$p$ in $G_{k}$ of length $n$ that start at the vertex $1$ and end at the
vertex $m$, where $1\leq m<k$, and let $\mathcal{H}_{n,k,m}$ be the Hilbert
subspace of $\mathcal{H}_{n,k}$ defined by the orthonormal basis $\left\{
\left\vert p\right\rangle :p\in\mathcal{P}_{n,k,m}\right\}  $. \ Then the
representation $\Phi:TL_{n}\left(  d\right)  \longrightarrow\mathbb{C}U\left(
\mathcal{H}_{n,k}\right)  $ splits into the direct sum of representations
\[
\Phi=%
{\displaystyle\bigoplus\limits_{m=1}^{k-1}}
\Phi^{(m)}\text{ \ ,}%
\]
where $\Phi^{(m)}:TL_{n}\left(  d\right)  \longrightarrow\mathbb{C}U\left(
\mathcal{H}_{n,k,m}\right)  $ is the representation arising from the
projection $%
{\displaystyle\bigoplus\limits_{m=1}^{k-1}}
\mathbb{C}U\left(  \mathcal{H}_{n,k,m}\right)  \longrightarrow\mathbb{C}%
U\left(  \mathcal{H}_{n,k,m}\right)  $. \ Hence, the image $\operatorname{Im}%
\left(  \Phi\right)  $ of the representation $\Phi:TL_{n}\left(  d\right)
\longrightarrow\mathbb{C}U\left(  \mathcal{H}_{n,k}\right)  $ lies in the
direct sum of the algebras $\mathbb{C}U\left(  \mathcal{H}_{n,k,m}\right)  $,
$1\leq m<k$, i.e.,
\[
\operatorname{Im}\left(  \Phi\right)  \subseteq%
{\displaystyle\bigoplus\limits_{m=1}^{k-1}}
\mathbb{C}U\left(  \mathcal{H}_{n,k,m}\right)  \text{ \ .}%
\]

\end{corollary}

\bigskip

The above corollary gives us the latitude of searching for a compatible trace
from among all the traces $\widetilde{Tr}:\operatorname{Im}\left(
\Phi\right)  \longrightarrow\mathbb{C}$ which are constructed by taking any
linear combination of the \textbf{standard traces} $\widetilde{Tr}%
_{m}:\mathbb{C}U\left(  \mathcal{H}_{n,k,m}\right)  \longrightarrow\mathbb{C}%
$. \ Of these traces, the desired compatible trace is found to be the one
given in the following theorem:

\bigskip

\begin{theorem}
Let $\lambda_{\ell}=\sin\left(  \pi\ell/k\right)  $ be the components of the
eigenvector $\overrightarrow{\lambda}$ given in section 6, and let $N=%
{\textstyle\sum\nolimits_{\ell=1}^{k-1}}
\lambda_{\ell}\dim\left(  \mathcal{H}_{n,k,\ell}\right)  $. \ Then the trace
$\widetilde{Tr}:\operatorname{Im}\left(  \Phi\right)  \longrightarrow
\mathbb{C}$ defined by%
\[
\widetilde{Tr}=\frac{1}{N}%
{\displaystyle\sum\limits_{\ell=1}^{k-1}}
\lambda_{\ell}\widetilde{Tr}_{\ell}\text{ \ ,}%
\]
is compatible with the Markov trace, i.e., $\widetilde{Tr}$ is a trace such
that the diagram
\[%
\begin{array}
[c]{lcl}%
TL_{n}\left(  d\right)  & \quad\overset{\Phi}{\longrightarrow}\quad &
\operatorname{Im}\left(  \Phi\right)  \subseteq%
{\displaystyle\bigoplus\limits_{\ell=1}^{k-1}}
\mathbb{C}U\left(  \mathcal{H}_{n,k,\ell}\right)  \subset\mathbb{C}U\left(
\mathcal{H}_{n,k}\right) \\
& Tr_{n}\searrow\quad & \qquad\quad\quad\ \ \downarrow\widetilde{Tr}\\
&  & \qquad\quad\quad\ \ \mathbb{C}%
\end{array}
\]
is commutative. \ In other words, $\widetilde{Tr}$ is a trace such that
$Tr_{n}=\widetilde{Tr}\circ\Phi$.
\end{theorem}

\bigskip

Since the Markov trace $Tr_{n}:TL_{n}\left(  d\right)  \longrightarrow
\mathbb{C}$ is the unique trace satisfying the following three conditions

\begin{itemize}
\item $Tr_{n}\left(  1\right)  =1$

\item $Tr_{n}\left(  XY\right)  =Tr_{n}\left(  YX\right)  $ for all $X$ and
$Y$ in $TL_{n}\left(  d\right)  $

\item If $X\in TL_{n}\left(  d\right)  $, then $Tr_{n+1}\left(  XE_{n}\right)
=\frac{1}{d}Tr_{n}\left(  X\right)  $
\end{itemize}

\noindent the proof of the above theorem consists simply in verifying that
$\widetilde{Tr}\circ\Phi:TL_{n}\left(  d\right)  \longrightarrow\mathbb{C}$
satisfies each of these conditions.

\bigskip

\section{Intermediate summary}

\bigskip

But where are we in regard to our objective of creating a quantum algorithm
for approximating the value of the Jones polynomial at a root of unity of the
form $e^{2\pi i/k}$, where $k$ is an arbitrary positive integer?

\bigskip

For a knot (or link) given by the closure $\beta^{Tr}$ of an $n$-stranded
braid $\beta$, we have seen that the Jones polynomial is given by the
expression
\[
V_{\beta^{Tr}}\left(  t\right)  =-A^{2Writhe\left(  \beta\right)  }%
d^{n-1}Tr_{n}\left(  \rho_{A}\left(  \beta\right)  \right)  \text{ ,}%
\]
where $d$ and $A$ are indeterminate complex numbers related by the equation
$d=-A^{2}-A^{-2}$, and where $t=A^{-4}$.

\bigskip

Setting $A=e^{-2\pi i/2k}$ (which implies $d=2\cos\left(  \pi/k\right)  $ and
$t=e^{2\pi i/k}$), we have the value of the Jones polynomial at $t=e^{2\pi
i/k}$ is given by
\[
V_{\beta^{Tr}}\left(  e^{2\pi i/k}\right)  =-A^{2Writhe\left(  \beta\right)
}d^{n-1}Tr_{n}\left(  \rho_{A}\left(  \beta\right)  \right)  \text{ \ .}%
\]
Since $-A^{2Writhe\left(  \beta\right)  }d^{n-1}$ is easily computed, the task
of determining $V_{\beta^{Tr}}\left(  e^{2\pi i/k}\right)  $ reduces to that
of evaluating the trace
\[
Tr_{n}\left(  \rho_{A}\left(  \beta\right)  \right)  \text{ \ .}%
\]

\bigskip

But from the previous two sections, we have
\[
Tr_{n}\left(  \rho_{A}\left(  \beta\right)  \right)  =\widetilde{Tr}%
_{n}\left[  \left(  \Phi\circ\rho_{A}\right)  \left(  \beta\right)  \right]
=\widetilde{Tr}_{n}\left[  \left(
{\displaystyle\bigoplus\limits_{m=1}^{k-1}}
\Phi^{(m)}\circ\rho_{A}\right)  \left(  \beta\right)  \right]  =\frac{1}{N}%
{\displaystyle\sum\limits_{m=1}^{k-1}}
\lambda_{m}Tr\left[  \left(  \Phi^{(m)}\circ\rho_{A}\right)  \left(
\beta\right)  \right]  \text{ ,}%
\]
where $Tr$ denotes the standard trace, where $\lambda_{m}=\sin\left(  \pi
m/k\right)  $, and where $N=\sum_{m=1}^{k-1}\lambda_{m}\dim\left(
\mathcal{H}_{n,k,m}\right)  $. \ Thus, our objective reduces to finding the
trace of each of the following $k-1$ unitary transformations (called
\textbf{global gates})
\[
U^{(m)}=\left(  \Phi^{(m)}\circ\rho_{A}\right)  \left(  \beta\right)  \text{,
\ \ }1\leq m<k
\]

\bigskip

If the knot (or link) is given by the closure of a braid $\beta$ defined by
the word
\[
\beta=%
{\displaystyle\prod\limits_{\ell=1}^{L}}
b_{j\left(  \ell\right)  }^{\epsilon\left(  \ell\right)  }=b_{j(1)}%
^{\epsilon(1)}b_{j(2)}^{\epsilon(2)}\cdots b_{j(L)}^{\epsilon(L)}\text{ \ ,}%
\]
where $b_{1},b_{2},\ldots,b_{n-1}$ are the generators of the braid group
$B_{n}$, and where $\epsilon\left(  i\right)  =\pm1$ for $i=1,2,\ldots,L$,
then each unitary transformation $U^{(m)}=\left(  \Phi^{(m)}\circ\rho
_{A}\right)  \left(  \beta\right)  $ is given by
\[
U^{(m)}=%
{\displaystyle\prod\limits_{\ell=1}^{L}}
\left(  U_{j\left(  \ell\right)  }^{(m)}\right)  ^{\epsilon\left(
\ell\right)  }\text{ ,}%
\]
where $U_{j}^{(m)}$ denotes the unitary transformation (called an
\textbf{intermediate gate})%
\[
U_{j}^{(m)}=\left(  \Phi^{(m)}\circ\rho_{A}\right)  \left(  b_{j}\right)
\text{ \ \ , \ }1\leq m<k\text{, \ }1\leq j<n\text{.}%
\]
Thus, the trace $Tr_{n}\left(  \rho_{A}\left(  \beta\right)  \right)  $ we
seek to approximate is given by the following expression
\[
Tr_{n}\left(  \rho_{A}\left(  \beta\right)  \right)  =\frac{1}{N}%
{\displaystyle\sum\limits_{m=1}^{k-1}}
\lambda_{m}Tr\left[  U^{(m)}\right]  =\frac{1}{N}%
{\displaystyle\sum\limits_{m=1}^{k-1}}
\lambda_{m}Tr\left[
{\displaystyle\prod\limits_{\ell=1}^{L}}
\left(  U_{j\left(  \ell\right)  }^{(m)}\right)  ^{\epsilon\left(
\ell\right)  }\right]  \text{ \ .}%
\]

\bigskip

We are finally in a position to describe the quantum algorithm found in
\cite{Aharonov1} for approximating the Jones polynomial $V_{\beta^{Tr}}\left(
t\right)  $ at $t=e^{2\pi i/k}$ as a quantum algorithm consisting of the
completion of two sequences of steps, called \textbf{phases}. \ The first is a
preliminary phase called the \textbf{compilation phase}. \ After completion,
the compilation phase is immediately followed by a second phase, called the
\textbf{execution phase}.

\bigskip

\section{The Compilation Phase}

\bigskip

The compilation phase is illustrated in figure \ 25.

\bigskip%

\begin{center}
\includegraphics[
natheight=7.499600in,
natwidth=9.999800in,
height=3.0277in,
width=4.0274in
]%
{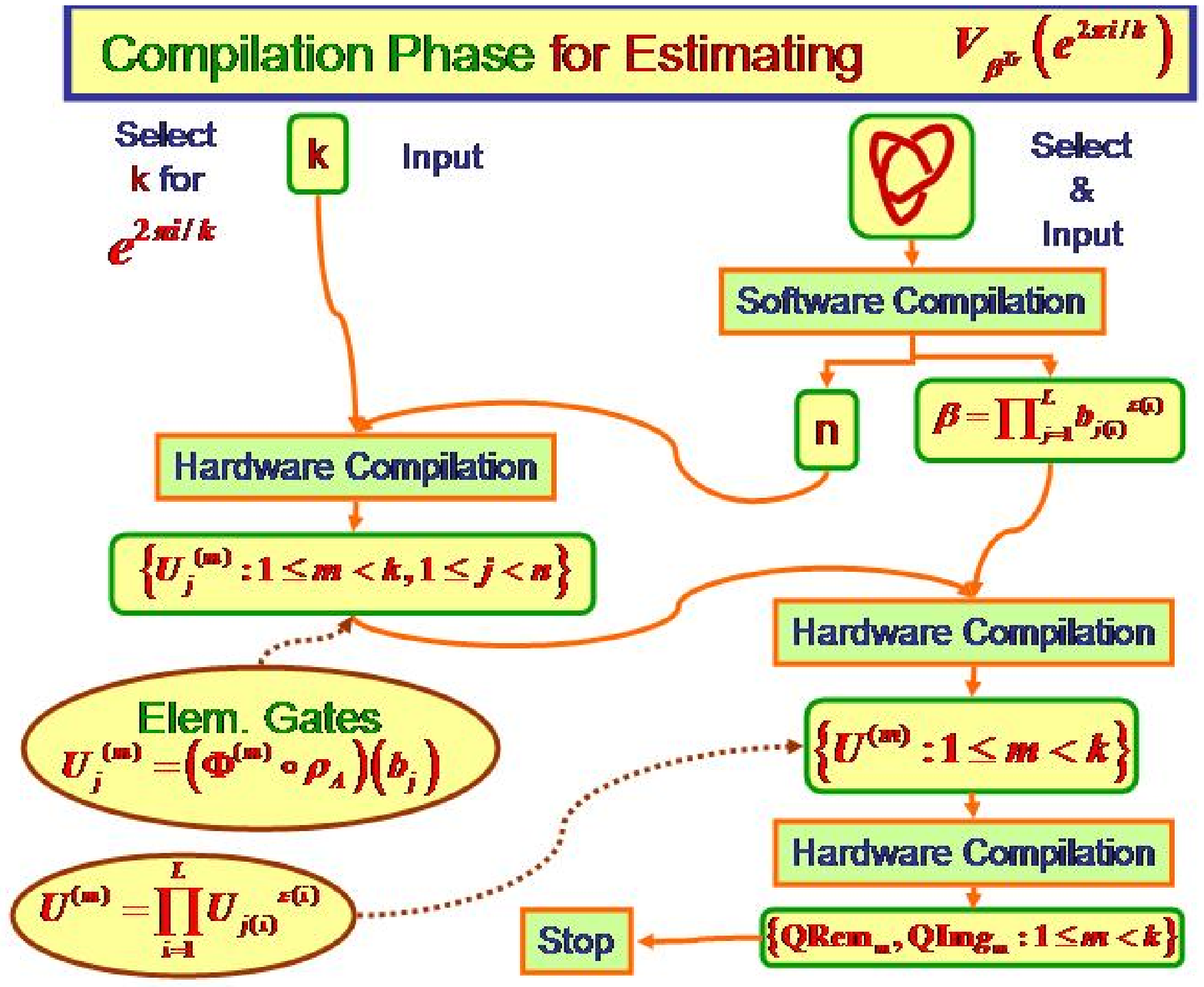}%
\\
\textbf{Figure 25. \ The compilation phase.}%
\end{center}

\bigskip

It consists of the following steps:

\bigskip

\noindent\textbf{Software Compilation.} On receiving a regular diagram of a
knot (or link) $K$ as input, the algorithm described by Alexander's theorem is
executed to produce a regular diagram of a braid $\beta$ whose closure gives
the knot (or link) $K$. \ (See Birman\cite{Birman1}.) \ An algorithm, called
\textbf{braid combing}, is then applied to the planar diagram of the braid
$\beta,$ producing as output a word
\[
\beta=%
{\displaystyle\prod\limits_{\ell=1}^{L}}
b_{j\left(  \ell\right)  }^{\epsilon\left(  \ell\right)  }=b_{j(1)}%
^{\epsilon(1)}b_{j(2)}^{\epsilon(2)}\cdots b_{j(L)}^{\epsilon(L)}%
\]
describing the braid, and also as a side effect, producing as output the
integer $n$ giving the number of strands in $\beta$. \ (Once again, see
Birman\cite{Birman1}.) \ \textit{This word can be thought of as a computer
program which will later be compiled into hardware.} \ 

\bigskip

\noindent\textbf{First Hardware Compilation.} \ Upon receiving as input the
integers $k$ and $n$, use the Kitaev-Solovay\cite{Nielsen1},\cite{Aharonov1}
theorem to implement (translate into hardware) good approximations of each
intermediate gate $U_{j}^{(m)}$, $1\leq j<n$, $1\leq m<k$ as a product of
polynomially many elementary gates. \ 

\bigskip

\noindent\textbf{Second Hardware Compilation.} \ Use the braid word
\[
\beta=%
{\displaystyle\prod\limits_{\ell=1}^{L}}
b_{j\left(  \ell\right)  }^{\epsilon\left(  \ell\right)  }%
\]
to implement (i.e., to physically construct) the global gates $U^{(m)}$ $1\leq
m<k$ from the intermediate gates $U_{j}^{(m)}$. \ In other words, construct
$U^{(m)}$ using the formula
\[
U^{(m)}=%
{\displaystyle\prod\limits_{\ell=1}^{L}}
\left(  U_{j\left(  \ell\right)  }^{(m)}\right)  ^{\epsilon\left(
\ell\right)  }\text{ \ .}%
\]

\bigskip

\noindent\textbf{Third Hardware Compilation.} \ For each $m$ ($1\leq m<k$),
construct from the global gate $U^{(m)}$ two quantum subroutines $QRe_{m}$ and
$QIm_{m}$, which upon input $\left\vert p\right\rangle $ \ ($p\in
\mathcal{P}_{n,k,m}$), output at random a `0' or a `1' according to the
following probability distributions
\[%
\begin{array}
[c]{l}%
QRe_{m}\left(  \left\vert p\right\rangle \right)  =\left\{
\begin{array}
[c]{ll}%
0 & \text{with probability }Prob\left(  0\right)  =\frac{1}{2}+\frac{1}%
{2}\operatorname{Re}\left\langle p|U^{(m)}|p\right\rangle \\
& \\
1 & \text{with probability }Prob\left(  1\right)  =\frac{1}{2}-\frac{1}%
{2}\operatorname{Re}\left\langle p|U^{(m)}|p\right\rangle
\end{array}
\right. \\
\\
QIm_{m}\left(  \left\vert p\right\rangle \right)  =\left\{
\begin{array}
[c]{ll}%
0 & \text{with probability }Prob\left(  0\right)  =\frac{1}{2}-\frac{1}%
{2}\operatorname{Im}\left\langle p|U^{(m)}|p\right\rangle \\
& \\
1 & \text{with probability }Prob\left(  1\right)  =\frac{1}{2}+\frac{1}%
{2}\operatorname{Im}\left\langle p|U^{(m)}|p\right\rangle
\end{array}
\right.
\end{array}
\]
where $\operatorname{Re}\left\langle p|U^{(m)}|p\right\rangle $ and
$\operatorname{Im}\left\langle p|U^{(m)}|p\right\rangle $ denote respectively
the real and the imaginary parts of the bracket $\left\langle p|U^{(m)}%
|p\right\rangle $. \ Thus, if $QRe_{m}\left(  \left\vert p\right\rangle
\right)  $ and $QIm_{m}\left(  \left\vert p\right\rangle \right)  $ are
repeated executed, then we obtain respectively approximations of the real and
imaginary parts of the bracket $\left\langle p|U^{(m)}|p\right\rangle $, as
displayed below:
\[
\left\{
\begin{array}
[c]{rr}%
\left(  \#0^{\prime}s-\#1^{\prime}s\right)  /\left(  \#0^{\prime}%
s+\#1^{\prime}s\right)  \approx\operatorname{Re}\left\langle p|U^{(m)}%
|p\right\rangle  & \text{for \ }QRe_{m}\left(  \left\vert p\right\rangle
\right) \\
& \\
\left(  \#1^{\prime}s-\#0^{\prime}s\right)  /\left(  \#0^{\prime}%
s+\#1^{\prime}s\right)  \approx\operatorname{Im}\left\langle p|U^{(m)}%
|p\right\rangle  & \text{for \ }QIm_{m}\left(  \left\vert p\right\rangle
\right)
\end{array}
\right.
\]

\bigskip

These two quantum subroutines $QRe_{m}$ and $QIm_{m}$ can be implemented using
what has now come to be known as the \textbf{Hadamard test}. \ A wiring
diagram defining the quantum subroutine $QRe_{m}\left(  \left\vert
p\right\rangle \right)  $ is given in figure 26. \ Input to $QRe_{m}$ consists
of the state $\left\vert p\right\rangle $ and an ancillary qubit in state
$\left\vert 0\right\rangle $. \ After execution of the the wiring diagram,
measurement of the ancillary qubit with respect to the standard basis will
produce the desired probability distribution. \ Similarly, Figure 27 defines
the quantum subroutine $QIm_{m}$.

\bigskip%

\begin{center}
\includegraphics[
trim=0.000000in 3.336572in 0.000000in 0.000000in,
natheight=7.499600in,
natwidth=9.999800in,
height=1.6924in,
width=4.0274in
]%
{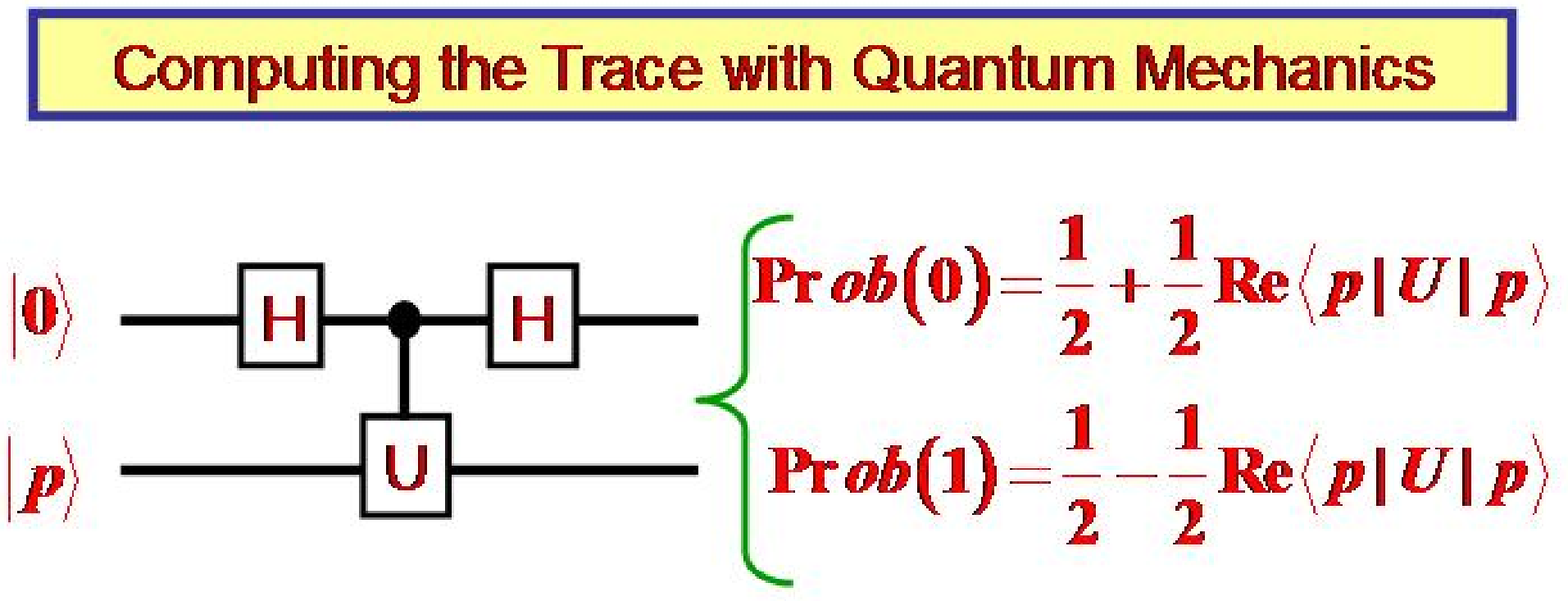}%
\\
\textbf{Figure 26. \ A wiring diagram describing the quantum subroutine
}$\mathbf{QRe}_{m}$\textbf{. Input consists of the state }$\left\vert
p\right\rangle $\textbf{ and an ancillarary qubit in state }$\left\vert
0\right\rangle $\textbf{. \ The Hadamard gate is denoted by }$H$.
\textbf{After execution of the first hadard gate, the ancillary qubit is used
to control the global gate }$U^{(m)}$\textbf{. Measurement of the ancillary
qubit after the execution of the second Hadamard gate produces the desired
probability distribution.}%
\end{center}
\begin{center}
\includegraphics[
trim=0.000000in 3.336572in 0.000000in 0.000000in,
natheight=7.499600in,
natwidth=9.999800in,
height=1.6924in,
width=4.0274in
]%
{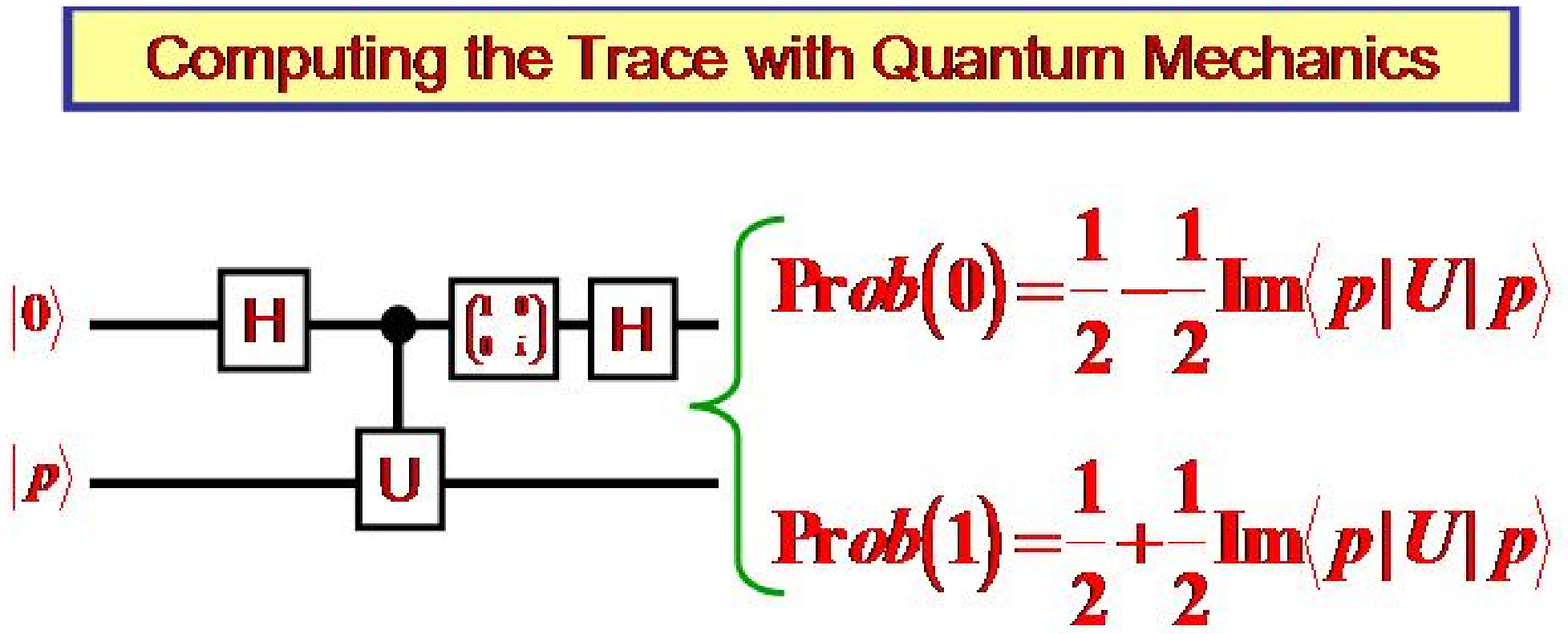}%
\\
\textbf{Figure 27. \ A wiring diagram describing the quantum subroutine
}$\mathbf{QIm}_{m}$\textbf{. This subroutine is the same as the quantum
subroutine }$\mathbf{QRe}_{m}$ \textbf{except for the additional single qubit
phase gate }$\left(
\begin{array}
[c]{rr}%
1 & 0\\
0 & i
\end{array}
\right)  $\textbf{.}%
\end{center}

\bigskip

\section{The Execution Phase}

\bigskip

The compilation phase is then followed by the execution phase as described by
the pseudocode given below:%

\[
\fbox{%
\begin{tabular}
[c]{l}\hline\hline
\hspace{0.75in}\textbf{Execution Phase for Estimating }$\mathbf{V}_{\beta
^{Tr}}\left(  e^{2\pi i/k}\right)  $\\\hline\hline
\\
\textsc{Algorithm} $AJK\left(  n,k\right)  $\\
\quad$Trace=0$\\
\quad\textsc{loop} \ $m=1\ldots k-1$\\
\quad\quad$Trace_{m}=0$\textsc{ \ and \ \ }$\lambda_{m}=\sin\left(  \pi
m/k\right)  $\\
\quad\quad\textsc{loop} \ $p\in\mathcal{P}_{n,k,m}$\\
\quad\quad\quad$Re=0$\textsc{ \ and \ }$Im=0$\\
\quad\quad\quad\textsc{loop} \ $Iteration=1\ldots NumberofIterations$\\
\quad\quad\quad\quad$RealBit=QRe_{m}\left(  p\right)  $\textsc{ \ and
\ \ }$ImgBit=QIm_{m}\left(  p\right)  $\\
\quad\quad\quad\quad$Re=Re+\left(  -1\right)  ^{RealBit}$\textsc{ \ \ and
\ \ }$Im=Im-\left(  -1\right)  ^{ImgBit}$\\
\quad\quad\quad\textsc{end} \ $Iteration$\textsc{-loop}\\
\quad\quad\quad$Trace_{m}=Trace_{m}+\left(  Re+i\ast Im\right)
/NumberofIterations$\\
\quad\quad\textsc{end} $p$\textsc{-loop}\\
\quad\quad$Trace=Trace+\lambda_{m}\ast Trace_{m}$\\
\quad\textsc{end} $m$\textsc{-loop}\\
\quad\textsc{output} $Trace$\\
\textsc{end Algorithm} $AJK$%
\end{tabular}
}%
\]

\bigskip

This phase consists of three nested loops. \ The innermost loop calls the
quantum subroutines $QRe_{m}$ and $QIm_{m}$. \ The parameter
$NumberofIterations$ is chosen according to the Chernoof-Hoeffding bound to
provide the desired accuracy for the approximation. \ If resources are
available, the outermost iteration loop can be replaced by a parallel implementation.

\bigskip

\section{Conclusion}

\bigskip

Indeed, much more could be said about the AJL quantum algorithm for the Jones
polynomial. \ But, for the time being, we will leave that task to one of our
future forthcoming papers on this subject.

\bigskip

\section{Acknowledgements}

\bigskip

This work is partially supported by the Defense Advanced Research Projects
Agency (DARPA) and Air Force Research Laboratory, Air Force Materiel Command,
USAF, under agreement number F30602-01-2-0522. The U.S. Government is
authorized to reproduce and distribute reprints for Governmental purposes
notwithstanding any copyright annotation thereon. \ This work also partially
supported by the Institute for Scientific Interchange (ISI), Torino, the
National Institute of Standards and Technology (NIST), the Mathematical
Sciences Research Institute (MSRI), the Isaac Newton Institute for
Mathematical Sciences, and the L-O-O-P fund.


\bigskip

\begin{thebibliography}{99}                                                                                               %


\bibitem {Aharonov1}Aharonov, Dorit, Vaughan Jones, and Zeph Landau, \textbf{A
polynomial quantum algorithm for approximating the Jones polynomial}, http://arxiv.org/abs/quant-ph/0511096

\bibitem {Artin1}Artin, E., \textbf{Theory of braids}, Annals of Mathematics,
\textbf{48}, (1947), 101-126.

\bibitem {Bernstein1}Bernstein, E., and U. Vazirani, \textbf{Quantum
complexity theory}, SIAM Journal of Computation, 26, 5, (1997), 1411-1473.

\bibitem {Birman1}Birman, J., \textbf{"Braids, links and mapping class
groups,"} Annals of Mathematical Studies, vol. 82, (1974).

\bibitem {Crowell1}Crowell, Richard H., and Ralph H. Fox,\textbf{ "Inroduction
to Knot Theory,"} Springer-Verlag, (1977).

\bibitem {Ewing1}Ewing, B. and K.C. Millett, \textbf{A load balanced algorithm
for the calculation of the polynomial knot and link invariants} in
\textbf{"The mathematical heritage of C. F. Gauss,"} 225--266, World Sci.
Publishing, River Edge, NJ, (1991).

\bibitem {Fox1}Fox, R.H., \textbf{A quick trip through knot theory}, in
\textbf{"Topology of 3-Manifolds and Related Topics,"} M.K. Fort, Jr., (Ed.),
Printice-Hall, (1962).

\bibitem {Freedman1}Freedman, M., \textbf{P/NP and the quantum field
computer}, Proc. Natl. Acad. Sci., USA, 95, (1998), 98--101

\bibitem {Freedman2}Freedman, M., A.Kitaev, M. Larsen, and Z. Wang,
\textbf{Topological quantum computation} in \textbf{"Mathematical challenges
of the 21st century (Los Angeles, CA, 2000),"} Bull. Amer. Math. Soc., 40 ,
no. 1, (2003), 31--38

\bibitem {Freedman3}Freedman, M.H., A. Kitaev, and Z. Wang, \textbf{Simulation
of topological field theories by quantum computers}, Commun. Math. Phys., 227,
(2002) 587-603.

\bibitem {Freedman4}Freedman, M.H., M. Larsen, and Z. Wang, A modular Functor
which is universal for quantum computation, Commun. Math. Phys., 227 no. 3,
(2002), 605-622.

\bibitem {Goodman1}Goodman, F.M., P. de la Harpe, and V.F.R. Jones,
\textbf{Coxeter graphs and towers of algebras}, Springer-Verlag, (1989).

\bibitem {Garnerone1}Garnerone, Silvano, Annalisa Marzuoli, and Mario Rasetti,
\textbf{Quantum automata, braid group and link polynomials}, http://arxiv.org/abs/quant-ph/0601169

\bibitem {Jaeger1}Jaeger, F., D.L. Vertigan, D.J.A. Welsh,\textbf{ On the
computational complexity of the Jones and Tutte polynomials}, Math. Proc.
Cambridge Philos. Soc., 108, no. 1, (1990), 35--53.

\bibitem {Jones1}Jones, V.F.R., \textbf{A polynomial invariant for knots via
von Neumann algebras}, Bull. Amer. Math. Soc., 12, no. 1, (1985), 103--111.

\bibitem {Jones2}Jones, V.F.R., \textbf{Index for subfactors}, Invent. Math 72
(1983), 1--25.

\bibitem {Jones3}Jones, V.F.R., \textbf{Braid groups, Hecke Algebras and type
II factors}, in \textbf{"Geometric methods in Operator Algebras,"} Pitman
Research Notes in Math., 123 (1986), 242--273.

\bibitem {Kauffman1}Kauffman, L., \textbf{State models and the Jones
polynomial}, Topology 26, (1987),395-407.

\bibitem {Kauffman2}Kauffman, Louis H., and Sostenes L. Lins,
\textbf{"Temperley-Lieb Recoupling Theory and Invariants of 3-Manifolds,"}
Princeton University Press, (1994).

\bibitem {Kauffman3}Kauffman, Louis H., \textbf{"Knots and Physics,"} (second
edition), World Scientific, (1993)

\bibitem {Kauffman4}Kauffman, Louis H., \textbf{Quantum computing and the
Jones polynomial} in \textbf{"Quantum computation and information,"} edited by
S.J. Lomonaco, Jr., AMS CONM/305, (2000), 101-137.

\bibitem {Kuperberg1}Kuperberg, G., \textbf{A subexponential-time quantum
algorithm for the dihedral hidden subgroup problem},
http://arxiv.org/abs/quant-ph/0302112 .

\bibitem {Lickorish1}Lickorish, W.B. Raymond, \textbf{"An Introduction to Knot
Theory,"} Springer-Verlag, (1997).

\bibitem {Lomonaco1}Lomonaco, Samuel J., Jr., \textbf{"Quantum Computation: A
Grand Mathematical Challenge for the Twenty-First Century and the
Millennium,"} AMS PSAPM 58, (2000).

\bibitem {Lomonaco2}Lomonaco, Samuel J., Jr., (ed.), \textbf{"Low Dimensional
Topology,"} AMS CONM/20, (1981).

\bibitem {Markov1}Markov, A., \textbf{Uber de freie Aquivalenz geschlossener
Zopfe}, Rossiiskaya Akademiya Nauk, Matematicheskii Sbornik, 1, (1935), 73--78.

\bibitem {Murasugi1}Murasugi, Kunio, \textbf{"Knot Theory and Its
Applications,"} Birkhauser, (1996).

\bibitem {Nielsen1}Neilsen, Michael A., and Isaac L. Chuang, \textbf{"Quantum
Computation and Quantum Information,"} Cambridge University Press, (2000).

\bibitem {Rolfsen1}Roldsen, Dale, \textbf{"Knots and Links,"} Publish or
Perish, Inc., (1976).

\bibitem {Vogel1}Vogel, P., \textbf{Representation of links by braids: A new
algorithm,} Comment. math. Helvetici, 65 (1) (1990), 104--113.

\bibitem {Welsh1}Welsh, D.J.A., \textbf{The Computational Complexity of Some
Classical Problems from Statistical Physics}, in \textbf{"Disorder in Physical
Systems,"} G.R. Grimmett and D.J.A. Welsh, eds., Clarendon Press, Oxford,
(1990), 307-321.

\bibitem {Welsh2}Welsh, D.J.A., "Complexity: Knots, Colourings and Counting,"
Cambridge University Press, (2000).

\bibitem {Witten1}Witten, Edward, \textbf{Quantum field theory and the Jones
polynomial}, Comm. Math. Phys., 121, no. 3, (1989), 351--399.

\bibitem {Wu1}Wu, F.Y., \textbf{Knot Theory and statistical mechanics}, Rev.
Mod. Phys. 64, No. 4., (1992).

\bibitem {Wocjan1}Wocjan, Pawel, and Jon Yard, \textbf{The Jones polynomial:
quantum algorithms and applications in quantum complexity theory}, http://arxiv.org/abs/quant-ph/0603069
\end{thebibliography}
\end{document}